\documentclass[fleqn,usenatbib]{mnras}

\usepackage{newtxtext,newtxmath}
\usepackage{verbatim}
\usepackage{orcidlink}  %ORCID LOGO

\usepackage[T1]{fontenc}

\DeclareRobustCommand{\VAN}[3]{#2}
\let\VANthebibliography\thebibliography
\def\thebibliography{\DeclareRobustCommand{\VAN}[3]{##3}\VANthebibliography}

\usepackage{graphicx}	% Including figure files
\usepackage{amsmath}	% Advanced maths commands
\usepackage{threeparttable} %Table captions
\usepackage{caption}
\usepackage{subcaption}
\makeatletter
\newcounter{parentsubcaption}
\newenvironment{subsubcaption}
 {\refstepcounter{sub\@captype}%
  \protected@edef\theparentsubcaption{\@nameuse{thesub\@captype}}%
  \ignorespaces
}{%
  \setcounter{sub\@captype}{\value{parentsubcaption}}%
  \ignorespacesafterend
}
\makeatother
\title[The rapid orbital decay of GX 301--2]{Investigating the orbital evolution of the eccentric HMXB GX 301--2 using long-term X-ray lightcurves}

\author[H. Manikantan et al.]{
Hemanth Manikantan,$^{1}$\thanks{E-mail: hemanthm@rri.res.in}${\orcidlink{0000-0001-9404-1601}}$
Manish Kumar,$^{1}\orcidlink{0009-0005-1889-8597}$
Biswajit Paul$^{1}$
and Vikram Rana$^{1}$${\orcidlink{0000-0003-1703-8796}}$
\\
$^{1}$Astronomy \& Astrophysics Department, Raman Research Institute, CV Raman Avenue, Sadashivanagar, Bangalore, India\\
}

\date{Accepted XXX. Received YYY; in original form ZZZ}

\pubyear{2023}

\begin{document}
\label{firstpage}
\pagerange{\pageref{firstpage}--\pageref{lastpage}}
\maketitle

\begin{abstract}
We report the orbital decay rate of the high mass X-ray binary GX 301--2 from an analysis of its long-term X-ray light curves and pulsed flux histories from \textit{CGRO}/BATSE, \textit{RXTE}/ASM, \textit{Swift}/BAT, \textit{Fermi}/GBM and \textit{MAXI} by timing the pre-periastron flares over a span of almost 30 years. The time of arrival of the pre-periastron flares exhibits an energy dependence (hard lag) and the orbital period decay was estimated after correcting for it. This method of orbital decay estimation is unaffected by the fluctuations in the spin rate of the X-ray pulsar associated with variations in the mass accretion rate. The resulting $\dot P_\textrm{orb}$ $=-(1.98\pm0.28)\times10^{-6}$ s s$^{-1}$ indicates a rapid evolution timescale of $|P_\textrm{orb}/\dot P_\textrm{orb}|\sim 0.6\times10^{5}$ yr, making it the high mass X-ray binary with the fastest orbital decay. Our estimate of $\dot P$\textsubscript{orb} is off by a factor of $\sim2$ from the previously reported value of $-(3.7\pm0.5)\times10^{-6}$ s s$^{-1}$ estimated from pulsar TOA analysis. We discuss various possible mechanisms that could drive this rapid orbital decay and also suggest that GX 301--2 is a prospective Thorne-\.{Z}ytkow candidate.
\end{abstract}


%%%%%%%%%%%%%%%%%%%%%%%%%%%%%%%%%%%%%%%%%%%%%%%%%%

%%%%%%%%%%%%%%%%% BODY OF PAPER %%%%%%%%%%%%%%%%%%

\section{Introduction}
Accreting High Mass X-ray Binary (HMXB) pulsars that host a rotating neutron star accreting matter from a companion star are hypothesised to be born from the supernova explosion of the more massive star in a preliminary binary stellar system hosting two relatively massive components ($>$12 M$_\odot$) (\citealt{tauris_vandenheuvel_2006}, and references therein). When the mass of the companion star is over 10 M$_\odot$, and it is of OB spectral type, they are called Supergiant HMXBs (SGXBs), which account for about one-third of the known HMXBs \citep{tauris_vandenheuvel_2006}. The binary orbit of SGXBs is postulated to evolve due to (i) tidal interactions, which also causes circularization of the eccentric orbit, (ii) mass transfer from companion to the neutron star by accretion, (iii) loss of mass from the binary by the stellar wind from the companion, and (iv) radiation by gravitational waves (\citealt{paul2011transient}, and references therein). The most accurate estimation of the orbital parameters and, thereby, the orbital evolution of accreting X-ray pulsars are obtained by measuring the time of arrival of the stable X-ray pulses from the X-ray pulsar. This technique is called the pulse time of arrival (TOA) analysis. The pulse TOA technique optimizes a parameter space comprising intrinsic pulse emission time stamps from the pulsar (accounting for inherent pulse period derivatives) and the binary orbit-induced arrival time delays in order to obtain the observed time of arrivals of each X-ray pulse \citep{nagase-intro}. Pulse timing analysis has been extensively used to accurately estimate the orbital evolution of SGXBs like Cen X--3, SMC X--1, LMC X--4, OAO 1657--415 and 4U 1538--52 \citep[see][and references therein]{paul_hmxb_orb_evolve_review}. Orbital decay (shrinking orbit) was observed in all the HMXBs hosting a pulsar, and the estimated decay time scale $|P_\textrm{orb}/\dot P_\textrm{orb}|$ varies from $\sim10^6$ yr in SMC X--1, Cen X--3, LMC X--4 and 4U 1538--52 to about $\sim10^7$ yr in OAO 1657--415 and 4U 1700--37 (Table~\ref{tab:hmxb_orbevolve}).

GX 301--2 is a rare galactic SGXB because of the unusually eccentric ($e\sim$0.47) binary orbit  \citep{sato1986-orb}, which is a peculiarity of Be-HMXBs (\citealt{paul2011transient}), and the only SGXB known to have a Hypergiant companion \citep{kaper_massloss_estimate_1995}. GX 301--2 is located $\sim5.3$ kpc away on the galactic plane and hosts a $\sim50$ M$_\odot$ Hypergiant stellar companion Wray 15-977 (BP Crucis) \citep{kaper_massloss_estimate_1995} and a NS in a $\sim41.5$ d long binary orbit \citep{sato1986-orb}. From the H$\alpha$ absorption profile in the optical spectrum, \cite{kaper_massloss_estimate_1995} estimated the mass loss rate from Wray 15-977 by the stellar wind to be $\lesssim10^{-5}$ M$_\odot$ yr$^{-1}$. A peculiar feature of GX 301--2 is its pre-periastron flaring nature, which is usually explained by enhanced accretion of matter from either a dense gas stream from the companion star (\citealt{gx301_gas_stream_haberl}, \citealt{LC_stream_model}) or an equatorial gas disc circumscribing Wray 15-977 \citep{Pravdo_2001}. Because of the pre-periastron flare, GX 301--2 exhibits variable X-ray intensity within each orbit, and the extent of the variation is energy dependent. The intensity varies by a factor of $\sim5$ in 4--10 keV and $\sim12$ in 15--50 keV. The wind of the companion star is clumpy \citep{mukherjee2003_gx301_clumpywind}, and it shows strong orbital phase dependent absorption column density and iron emission line \citep{islam2014_gx301_maxi,manikantan_gx301_maxi}. The pre-periastron flaring nature and binary ephemeris of GX 301--2 were first estimated by \cite{sato1986-orb} by pulse TOA analysis from \textit{SAS--3}, \textit{Hakucho}, and \textit{Ariel--V} observations. A similar analysis was performed by \cite{koh1997-orbit} on the \textit{CGRO}/BATSE data, and the reported orbital elements were consistent with \cite{sato1986-orb}. However, the orbital solution estimated by \cite{sato1986-orb} and \cite{koh1997-orbit} did not show any evidence of the decay of the orbital period. Evidence for orbital decay of the binary with $\dot P_\textrm{orb} =-(3.7\pm0.5)\times10^{-6}$ s s$^{-1}$ was later discovered by \cite{doroshenkov2010-orb} using pulse TOA analysis from a long \textit{INTEGRAL} observation (covering about 60\% of a binary orbit), under the assumption of a constant spin-up/down rate ($\dot P$\textsubscript{spin}) of the X-ray pulsar. This is the smallest orbital decay timescale observed in any HMXB. However, the large luminosity change of GX 301--2 along its orbital phase is most likely due to a variable mass accretion rate, and an important implication of the variable luminosity of GX 301--2 within each orbit is its effect on the spin-up rate of the pulsar. The spin-up rate of GX 301--2 is known to be correlated with the X-ray luminosity \citep{koh1997-orbit}. Previous estimations of orbital parameters and orbital evolution (\citealt{koh1997-orbit}, \citealt{doroshenkov2010-orb}), however, did not consider a luminosity-dependent period derivative \citep{2020monkkoen}.

In this work, we use the long-term X-ray lightcurves of GX 301--2 available from the X-ray All-sky monitors \textit{RXTE}/ASM, \textit{Swift}/BAT and \textit{MAXI} and the pulsed flux histories available from \textit{CGRO}/BATSE and \textit{Fermi}/GBM to investigate the orbital decay, which has previously been reported from pulse TOA analysis. Instead of the pulsar time stamps, which are used in the pulse TOA analysis, we make use of the similarity in the shapes of recurring orbital intensity profiles and the timing signature of the recurring pre-periastron flare peaks of GX 301--2. Assuming the orbital intensity profile of GX 301--2 to preserve an overall shape over the long term, epoch folding the long-term lightcurves and pulsed-flux histories could be used to estimate the orbital period and period derivative. Assuming that the physical mechanism responsible for the pre-periastron flares remains stable over the long term, we also utilize the variations in the arrival times of pre-periastron flares over an extended period to estimate the rate of change of the orbital period.
 
\section{Instrument and Observations}\label{sec:inst_obs}
The orbital period of GX 301--2 is relatively long, spanning 41.5 days (3586 ks), which makes conducting pointed observations throughout the entire orbit of GX 301--2 infeasible. However, being one of the brightest sources in the X-ray sky, GX 301--2 is monitored by all of the X-ray all-sky monitor observatories. The long-term lightcurves or pulsed flux histories from these observatories are available for over three decades.

The Burst and Transient Source Alert (BATSE) instrument onboard the \textit{Comption Gamma Ray Observatory (CGRO)} \citep{batse_meegan} consisted of eight inorganic NaI-based Scintillation detectors detecting hard X-ray photons from different parts of the sky in 20 keV -- 2 MeV. \textit{CGRO}/BATSE was operational from 1991 to 2000. The pulse periods of several X-ray pulsars were measured by epoch folding technique, and their pulse period and pulsed flux histories are available for download at the BATSE Pulsars webpage\footnote{\url{https://gammaray.nsstc.nasa.gov/batse/pulsar/}}.

The All-sky monitor (ASM) onboard the \textit{Rossi X-ray Timing Explorer} (RXTE) (\citealt{asm_levine}; \citealt{rxte_jahoda}) consisted of three position-sensitive Xenon proportional counters coupled to three coded-aperture masks respectively, and it operated in the 1.5--12 keV energy band. It had a total collecting area of 90 cm$^2$ and covered almost 80\% of the entire sky during each 90 min orbit, and it provided continuous data coverage of bright X-ray sources from 1996 to 2011.

The Burst Alert Telescope (BAT) onboard the Neil Gehrels Swift Observatory (\citealt{bat_2005}; \citealt{Swift_Gehrels_2004}) is a hard X-ray all-sky monitor operating in the 15--50 keV band. BAT consists of Cadmium Zinc Telluride (CZT) detectors (total detector area of about 5200 cm$^2$) coupled to a two-dimensional coded-aperture mask. This facilitates imaging of the X-ray sky with a large instantaneous field of view of 1.4 std. \textit{Swift}/BAT has been operational since 2004.

The Gamma Burst Monitor (GBM) onboard the \textit{Fermi Gamma-ray Space Telescope} is a hard X-ray monitor operating in 8 keV to 40 MeV. It consists of 12 Thallium activated Sodium Iodide (NaI(Tl)) scintillation detectors operating in 8 keV -- 1 MeV range and two Bismuth Gemanate (BGO) scintillation detectors operating in 200 keV - 40 MeV range. The GBM Accreting Pulsars Program (GAPP) provides the pulsed flux histories of bright X-ray pulsars (See \citealt{2malacaria_gapp} for a review). It is operational since 2008.

The Gas Slit Camera (GSC) onboard the \textit{Monitor of All-sky X-ray Image} (\textit{MAXI})  observatory (\citealt{gsc_maxi}; \citealt{maxi}) is an  All-sky monitor onboard the International Space Station (ISS) operating in the range 2--30 keV. GSC comprises twelve large-area position-sensitive proportional counters, each coupled to a slit-slat collimator. They have an instantaneous FOV of $160^\circ\times3^\circ$ and scan the whole sky during each orbit of the ISS. The narrow FOV and position-sensitive proportional counters facilitate imaging of the X-ray sky. The long-term lightcurves of X-ray sources from \textit{MAXI} are available since 2008.

We downloaded the orbit-by-orbit (dwell) long-term lightcurves from \textit{RXTE}/ASM (1.5--12 keV), \textit{Swift}/BAT (15--50 keV) and \textit{MAXI} (2--4, 4--10, 10--20 keV). The dwell lightcurves have a bin size of about 90 minutes (0.0625 days). However, the pulsed flux histories from \textit{CGRO}/BATSE and \textit{Fermi}/GBM were available with a bin size of 1 day and 2 days, respectively. The \textit{Swift}/BAT lightcurve was screened such that the data points having a value of error greater than 500 times the lowest error were excluded from the analysis.

\section{Analysis}
We performed three independent analyses to search for the orbital period decay in GX 301--2. In the first approach, epoch folding search \citep{epochfold_leahy_1987} was run on each long-term lightcurve without $\dot P_\textrm{orb}$, and the slope of the best-fitting straight line on the best periods derived from each of them as a function of time was estimated. In the second approach, epoch folding search was run on each long-term lightcurve for a prospective range of $\dot P_\textrm{orb}$ from $-3\times10^{-5}$ to $+3\times10^{-5}$ s s\textsuperscript{-1} to check if there is improved detection of periodicity corresponding to any $\dot P_\textrm{orb}$. This would indicate the presence of any period evolution in the long-term lightcurves. In the third approach, we used the times of the periodic pre-periastron flares to estimate the orbital period decay. The first two approaches depend on the long-term consistency of the orbital intensity profile of GX 301--2, which is dominated by the pre-periastron flare. The second approach depends on precisely locating the peak of the pre-periastron flares and the long-term stability of the time of arrival of pre-periastron flares. This means that the most significant factor affecting all three analyses is the accuracy of the shape of the flare. Since the flare is about 2 days long, the lightcurves used for analysis should preferably have a finer time resolution to construct the shape of the flare accurately. For this purpose, the 0.0625 d bin size dwell lightcurves were used for the analysis. However, the pulsed flux histories from BATSE and GBM were only available with a bin size of 1 d and 2 d, respectively, impacting the estimation accuracy from these two lightcurves.

\subsection{Epoch folding search}
\begin{figure}
    \centering
    \includegraphics[width=\columnwidth]{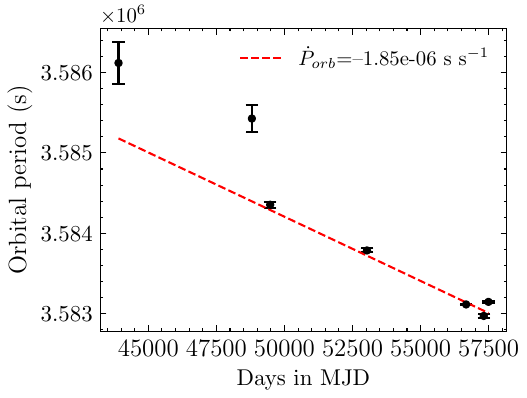}
    \caption{Estimates of the orbital periods from the long-term lightcurves and pulsed flux histories. $\dot P_\textrm{orb}$ from fitting a linear model is $-(1.85\pm0.34)\times10^{-6}$ s s\textsuperscript{-1}.}
    \label{fig:ltlc-efsearch}
\end{figure}

\begin{figure}
    \centering
    \includegraphics[width=\columnwidth]{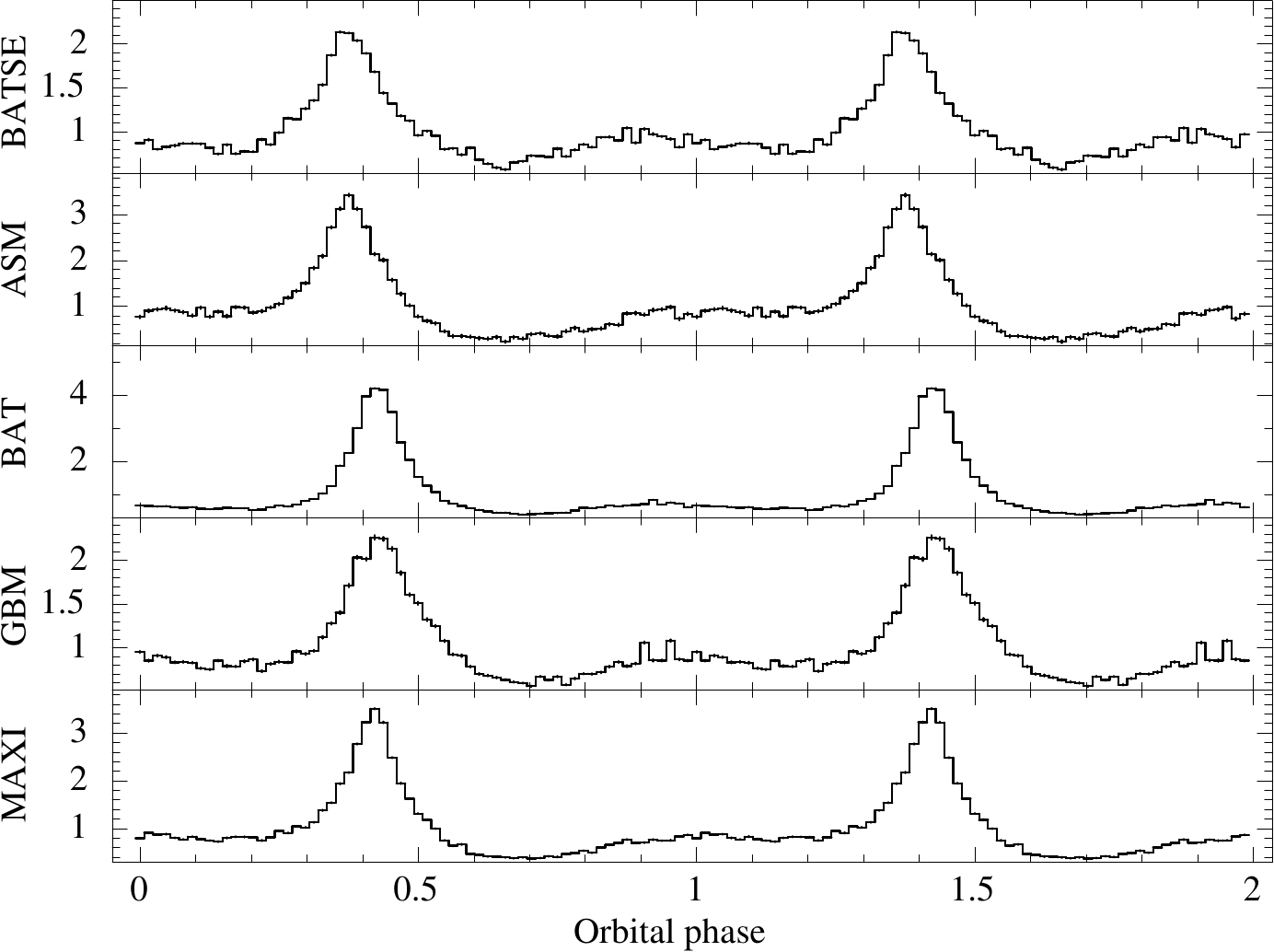}
    \caption{The orbital profiles of GX 301-2 obtained by folding the long-term lightcurves. The y-axis has units of cts s$^{-1}$ normalized by the average source count rate (normalized intensity). The lightcurves were folded at an epoch MJD 48370.5 corresponding to the beginning of BATSE lightcurve with the orbital periods derived from respective lightcurves (Table~\ref{tab:ltlc-efsearch}).}
    \label{fig:ltlc_orbit_profiles}
\end{figure}

We ran the epoch folding search over the entire duration of each of the three long-term lightcurves and two pulsed flux histories mentioned in Section~\ref{sec:inst_obs} (Also see Table~\ref{tab:overlap_lc_2} and Fig.~\ref{fig:ltlc}) using the {\small HEASOFT} tool \textit{efsearch}\footnote{\url{https://heasarc.gsfc.nasa.gov/ftools/fhelp/efsearch.txt}}. We searched for periods in the vicinity of 3583780 s (41.5 d), which is the known binary orbital period. For estimating the error in the best period returned by \textit{efsearch} in a lightcurve, we simulated 1000 instances of that particular lightcurve and ran \textit{efsearch} on each one of them, and the variance of the distribution of the best periods returned from 1000 lightcurves was used to estimate the $1\sigma$ error in the period (See Appendix~\ref{appendix:errorefsP} for details). The best period returned from each lightcurve was then assigned to the middle of the respective lightcurve duration and then plotted (Fig~\ref{fig:ltlc-efsearch}, Table~\ref{tab:ltlc-efsearch}). There is a clear trend of decreasing period, and a linear fit returns a best-fit orbital decay rate of $-(1.85\pm0.34)\times10^{-6}$ s s\textsuperscript{-1}. The orbital profiles obtained by folding each lightcurve with the respective orbital periods obtained from \textit{efsearch} are shown in Fig.~\ref{fig:ltlc_orbit_profiles}.

\begin{table*}
    \centering
    \caption{Estimates of the orbital period from the long term All sky monitor daily lightcurves and pulsed flux history lightcurves with associated $1\sigma$ error bars.}
    \label{tab:ltlc-efsearch}
    \begin{tabular}{cccc}
        \hline
        \hline
        Observatory &Epoch (MJD) &Best period &Reference\\
        \hline
         \textit{SAS-3}, \textit{Hakucho}, \textit{Ariel-5} $^\ddagger$ &43906.06  &$3586291.2\pm604.8$ &\cite{sato1986-orb}\\
         \textit{INTEGRAL} $^\ddagger$ &43906.06 &$3586118.4\pm259.2$ &\cite{doroshenkov2010-orb}\\
         \textit{CGRO}/BATSE $^\ddagger$ &48802.79 &$3585427.2\pm172.8$ &\cite{koh1997-orbit}\\
          \textit{CGRO}/BATSE &49475.00 &$3584355.24\pm36.34$ &This work.\\
         \textit{RXTE}/ASM$^\dagger$&53030.00 &$3583787.19\pm24.95$ &This work.\\
         \textit{Swift}/BAT &56671.32 &$3583115.88\pm3.92$ &This work.\\
         \textit{Fermi}/GBM &57318.77 &$3582973.00\pm22.67$ &This work.\\
         \textit{MAXI} &57489.84 &$3583147.78\pm9.45$ &This work.\\
         \hline
    \end{tabular}
    \begin{tablenotes}
        \item $^\dagger$ 1.5--12 keV band.
        \item $^\ddagger$ From pulse TOA analysis.
        \item The long-term lightcurves from \textit{RXTE}/ASM, \textit{Swift}/BAT and \textit{MAXI} have bin size of 0.0625 d. The pulsed flux history from \textit{CGRO}/BATSE and \textit{Fermi}/GBM have bin sizes of 1.0 d and 2.0 days, respectively.
    \end{tablenotes}
\end{table*}

\subsection{Epoch folding search with a period derivative}
\begin{figure*}
\centering
\begin{subsubcaption}
\begin{subfigure}{0.498\textwidth}
\includegraphics[width=\linewidth]{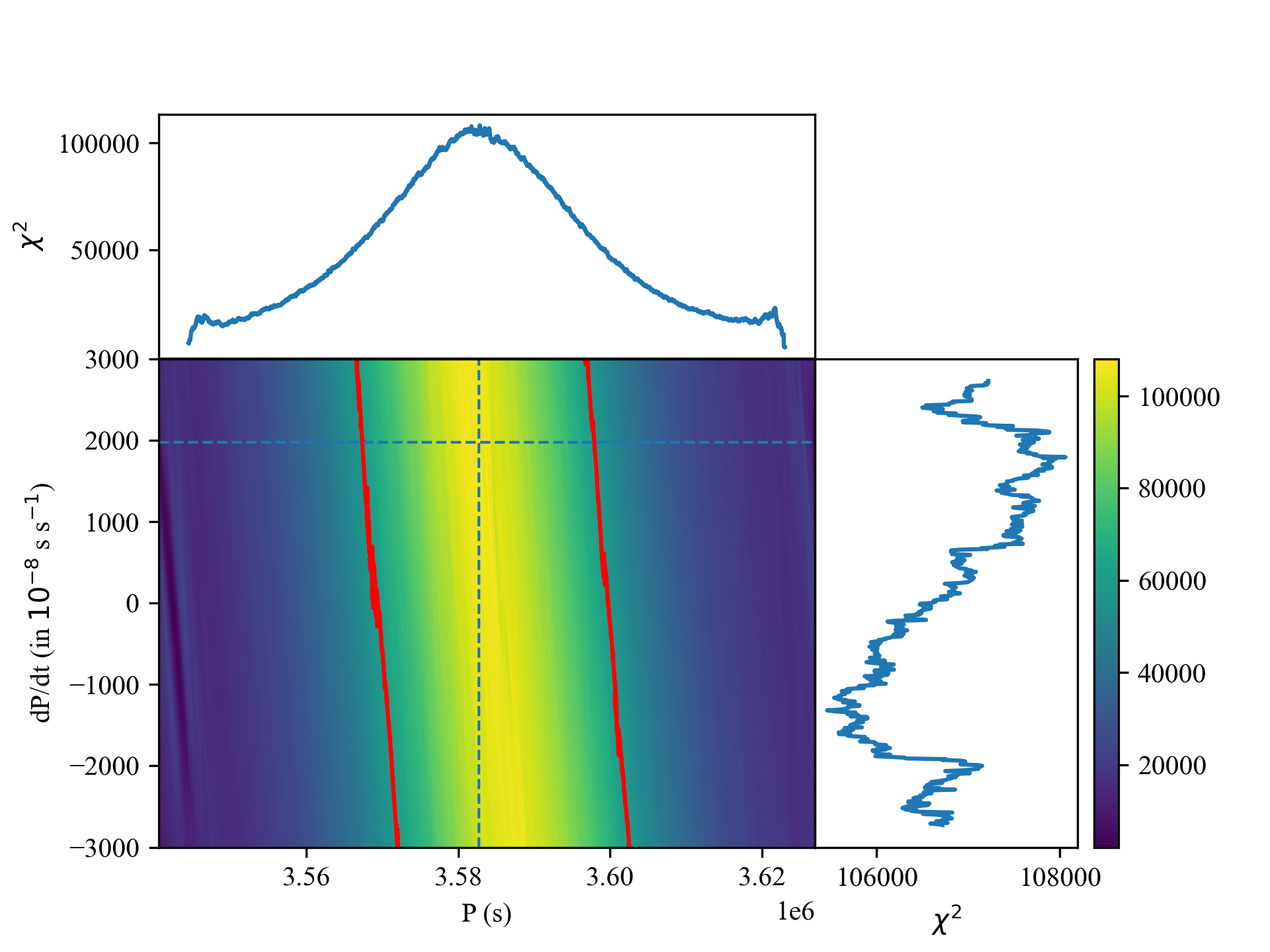}
\caption{\textit{CGRO}/BATSE}\label{subfig:pdot_batse}
\end{subfigure}
\begin{subfigure}{0.498\textwidth}
\includegraphics[width=\linewidth]{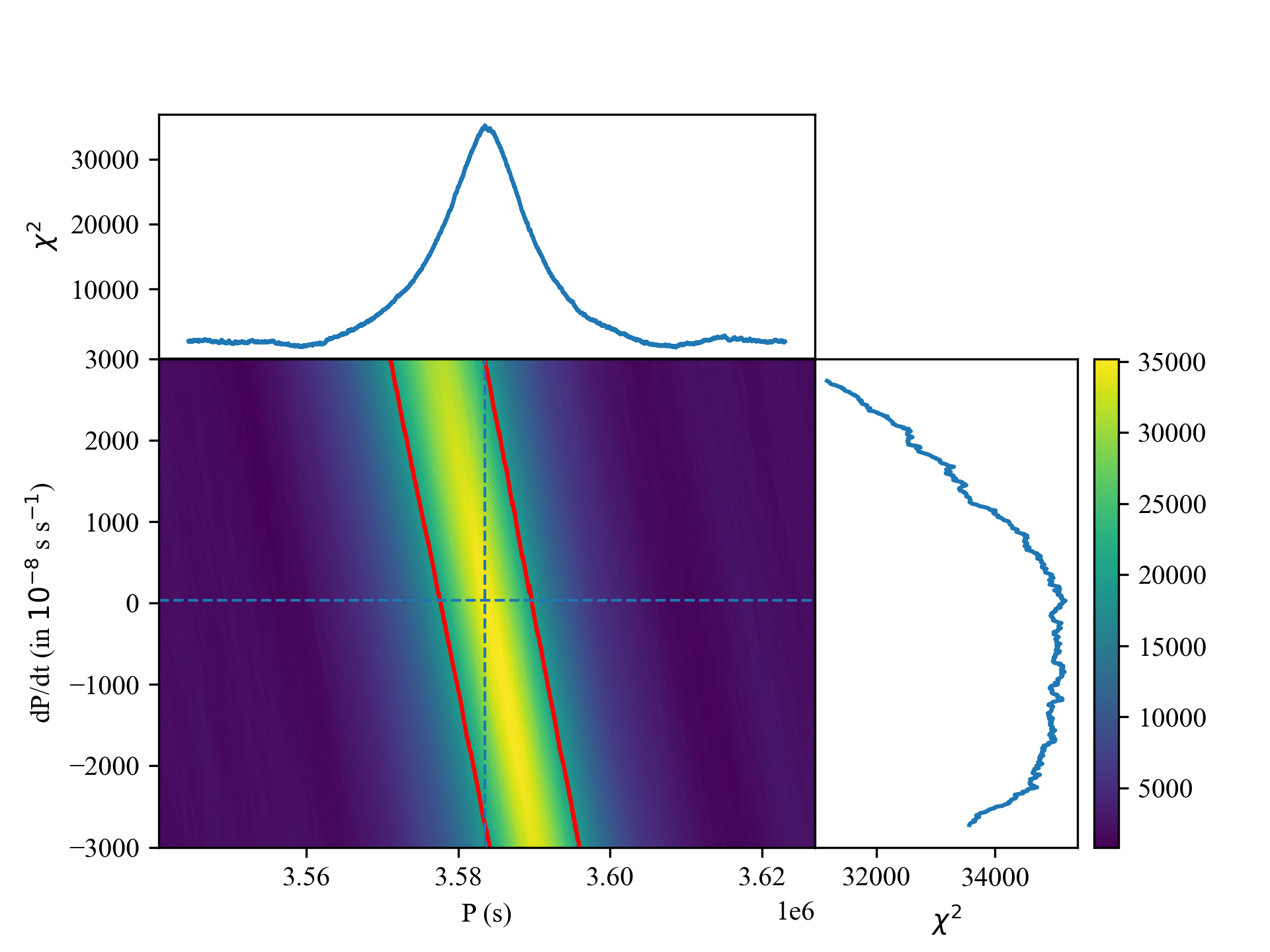}
\caption{\textit{RXTE}/ASM}\label{subfig:pdot_asm}
\end{subfigure}
\begin{subfigure}{0.498\textwidth}
\includegraphics[width=\linewidth]{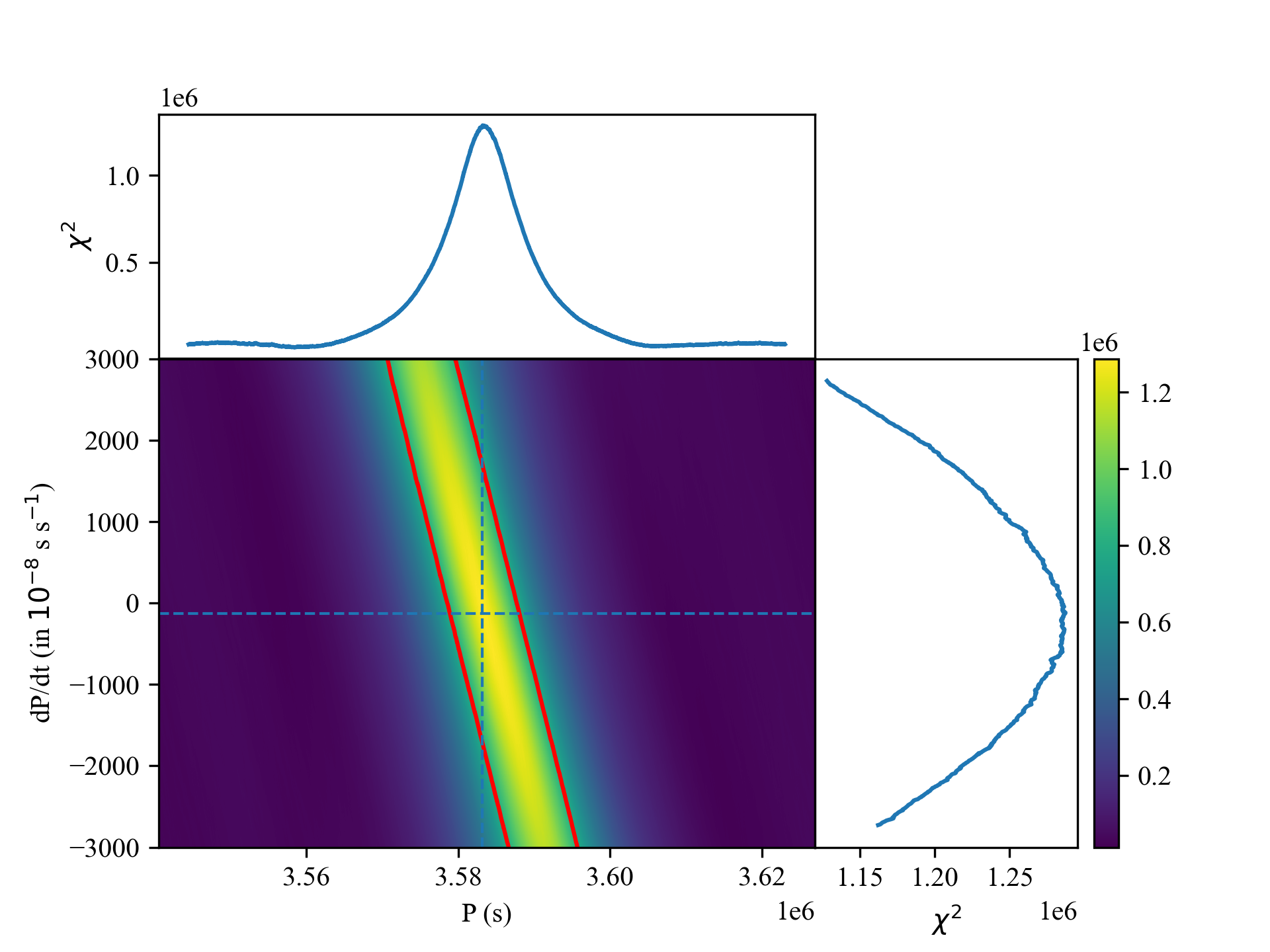}
\caption{\textit{Swift}/BAT}\label{subfig:pdot_bat}
\end{subfigure}
\begin{subfigure}{0.498\textwidth}
\includegraphics[width=\linewidth]{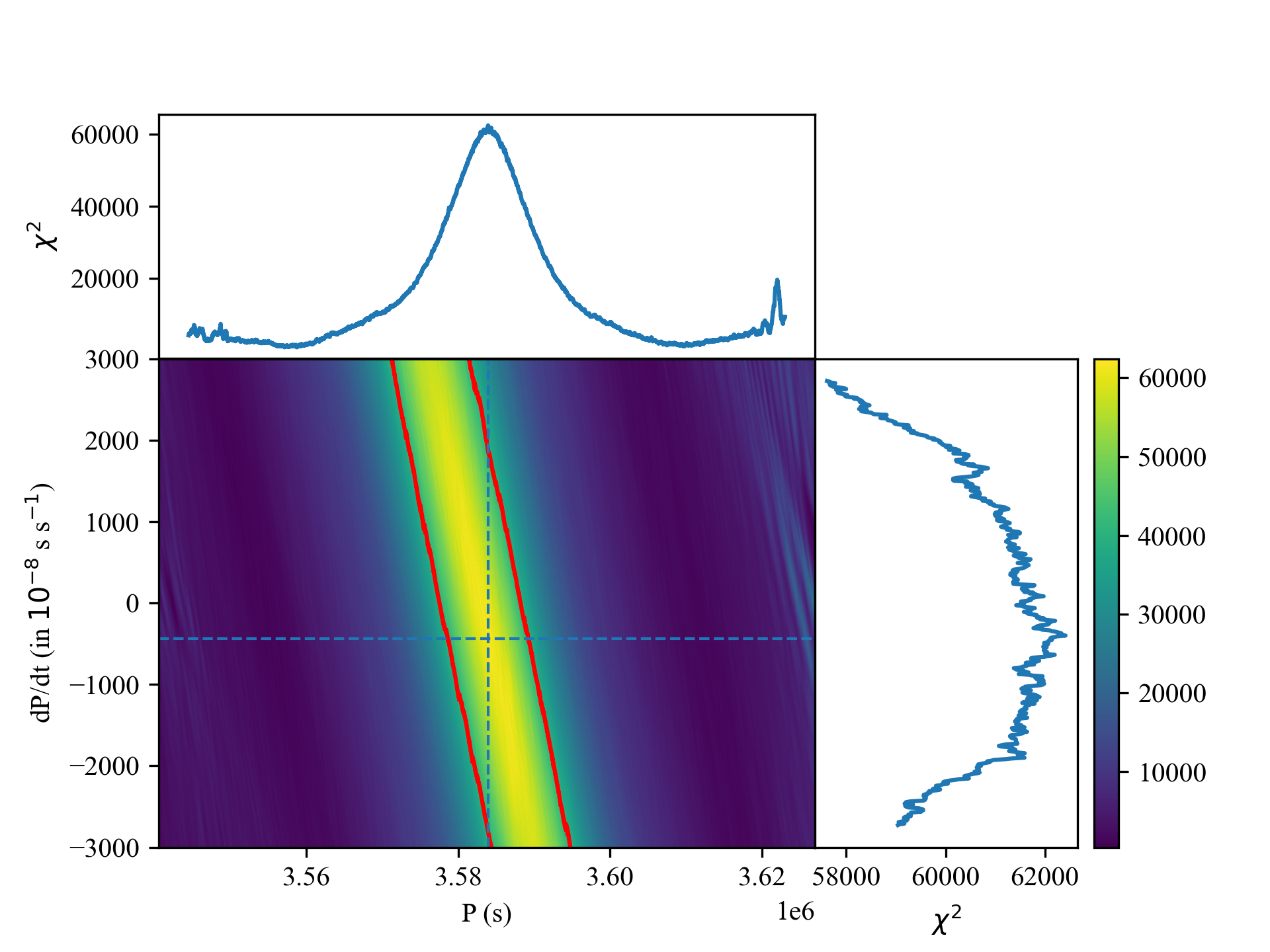}
\caption{\textit{Fermi}/GBM}\label{subfig:pdot_gbm}
\end{subfigure}
\begin{subfigure}{0.498\textwidth}
\includegraphics[width=\linewidth]{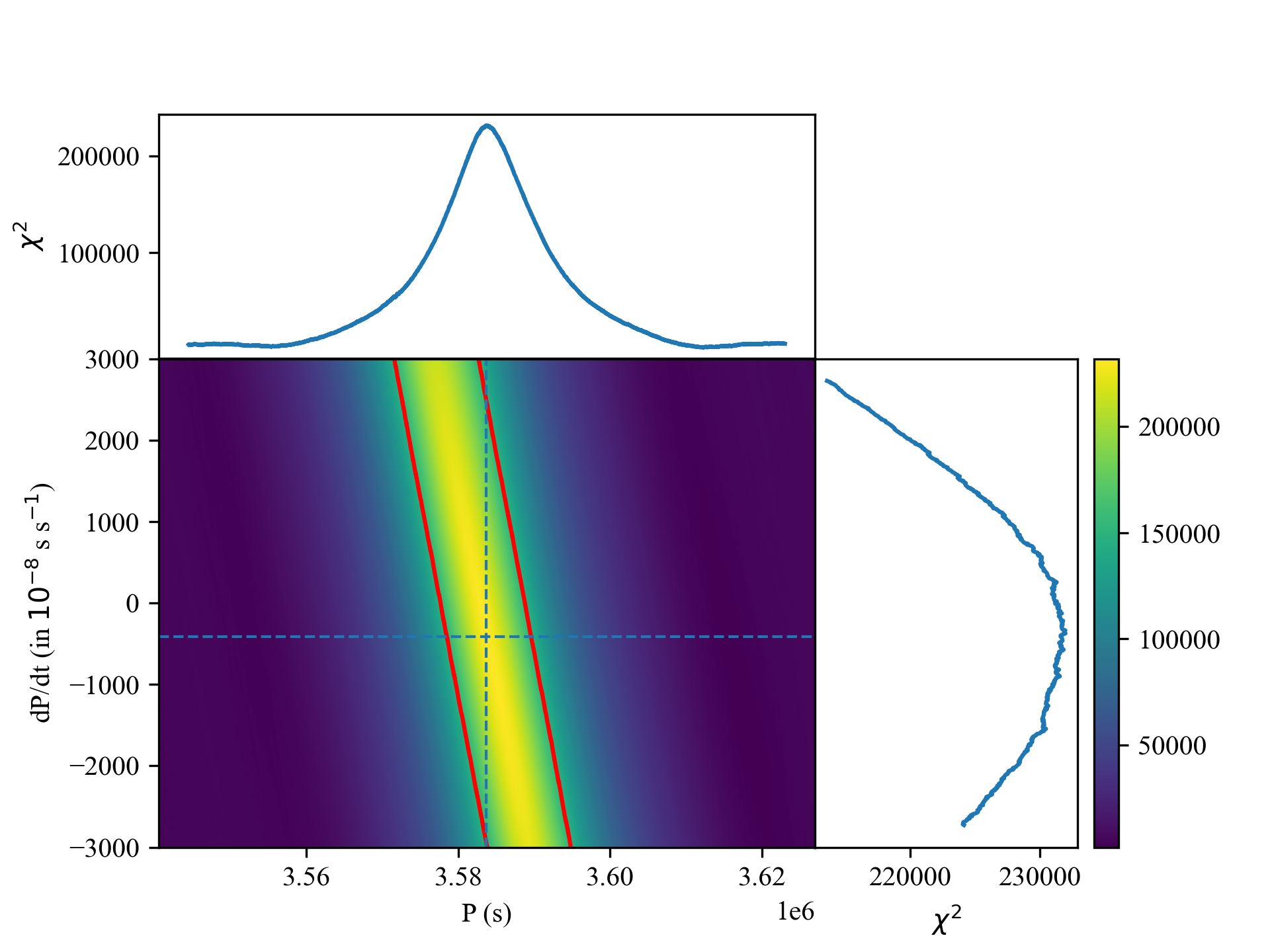}
\caption{\textit{MAXI}}\label{subfig:pdot_maxi}
\end{subfigure}
\end{subsubcaption}

\caption{Figure shows the results of \textit{efsearch} for different $\dot P$ values, when run on three long-term light curves and two pulsed flux histories. The epoch used for the period search for each lightcurve is the start of the respective lightcurve. Each plot has three panels, and the middle panel has $\chi^2-P$ horizontal plots stacked vertically for different values of $\dot P$, with $\chi^2$ colour coded. The top panel shows the $\chi^2-P$ plot returned for the best $\dot P$ ($\chi^2-P$ with the highest $\chi^2$\textsubscript{peak}), and the right panel shows the $\chi^2$\textsubscript{peak} obtained from each \textit{efsearch} run with a particular $\dot P$. The pair of horizontal and vertical dashed lines in the middle panel denotes ($P,\dot P$) corresponding to the highest $\chi^2$\textsubscript{peak} along each axes. \textit{Swift}/BAT, \textit{MAXI} and \textit{Fermi}/GBM show the presence of a secular $\dot P$\textsubscript{orb} of the order of $10^{-6}$ s s$^{-1}$, and \textit{RXTE}/ASM is consistent with such a value, but such a trend is not evident in \textit{CGRO}/BATSE.}
\label{fig:p-pdot1}
\end{figure*}

To search for the presence of such an orbital period decay within the duration of each lightcurve, we ran \textit{efsearch} in a range of sample period derivatives ranging from $-3\times10^{-5}$ to $+3\times10^{-5}$ s s\textsuperscript{-1} in each of the lightcurves. The results are shown in Fig.~\ref{fig:p-pdot1}. \textit{Swift}/BAT, \textit{Fermi}/GBM and \textit{MAXI} clearly show the presence of an orbital decay rate of around $-10^{-6}$ s s$^{-1}$  and \textit{RXTE}/ASM is consistent with this value (See the caption of Fig.~\ref{fig:p-pdot1}). However, such an orbital decay is not detected with \textit{CGRO}/BATSE.

\subsection{O--C curves using the pre-periastron flares}\label{subsec:o-c}
The recurring pre-periastron flares at regular intervals are a peculiarity of GX 301--2, and the time stamps ($T$\textsubscript{flare}) of pre-periastron flare peaks are useful markers to track the evolution of the binary orbital period. For a stable binary orbital period without temporal evolution, if the timestamp of flare in 0\textsuperscript{th} orbit ($T_0$) is known, the time stamp of flare in $n$\textsuperscript{th} orbit will follow the linear function $T_0+nP$\textsubscript{orb}. Any deviation from linearity in the observed time stamps of the flares indicates orbital period evolution. The difference between the observed and computed time of flares as a function of orbit cycle number is called the O--C curve. This technique was utilized for the estimation of orbital evolution by monitoring the time of arrivals of minima in the orbital intensity profile of Cyg X--3 (\citealt{singh_cygx3}) and by tracking the mid-eclipse times of eclipsing binaries Cen X--3, SMC X--1 \citep{cenx3_orbPdot_raichur_rxte} and LMC X--4 \citep{lmcx4_orbPdot_Naik_2004}. We use the same technique, with the orbital-intensity minima or mid-eclipse time substituted by pre-periastron flare peak (essentially the orbital-intensity maxima).

Since the photon statistics do not allow an accurate estimation of flare times for every orbital cycle from the long-term lightcurves, we constructed a representative flare peak time for short-duration segments of the long-term lightcurves. We divided each of the five lightcurves into three segments of equal duration and determined a representative time of arrival of the flare in each of those time segments. The time of arrival of the flare on $n$\textsuperscript{th} orbital cycle can be expressed as a Taylor polynomial function of $n$:
\begin{equation}
    T_n  = T_0 + \frac{n}{1!} P_\textrm{orb} + \frac{n^2}{2!} P_\textrm{orb} \dot P_\textrm{orb} + \ldots\label{eq:en}
\end{equation}

$T_n$ is the time stamp of the n\textsuperscript{th} pre-periastron flare peak, $T_0$ is the time stamp of the reference pre-periastron flare peak, $P_\textrm{orb}$ is the orbital period, and $\dot P_\textrm{orb}$ is the rate of change of orbital period derivative. Assuming $\dot P$\textsubscript{orb} is present and ignoring the higher order derivatives, equation~\ref{eq:en} can be used to verify the presence and get an estimate of $\dot P_\textrm{orb}$ if it exists (See \citealt{klis1984orbital_quadratic}; \citealt{cenx3_orbPdot_raichur_rxte}).

\begin{figure}
    \centering
    \includegraphics[width=\columnwidth]{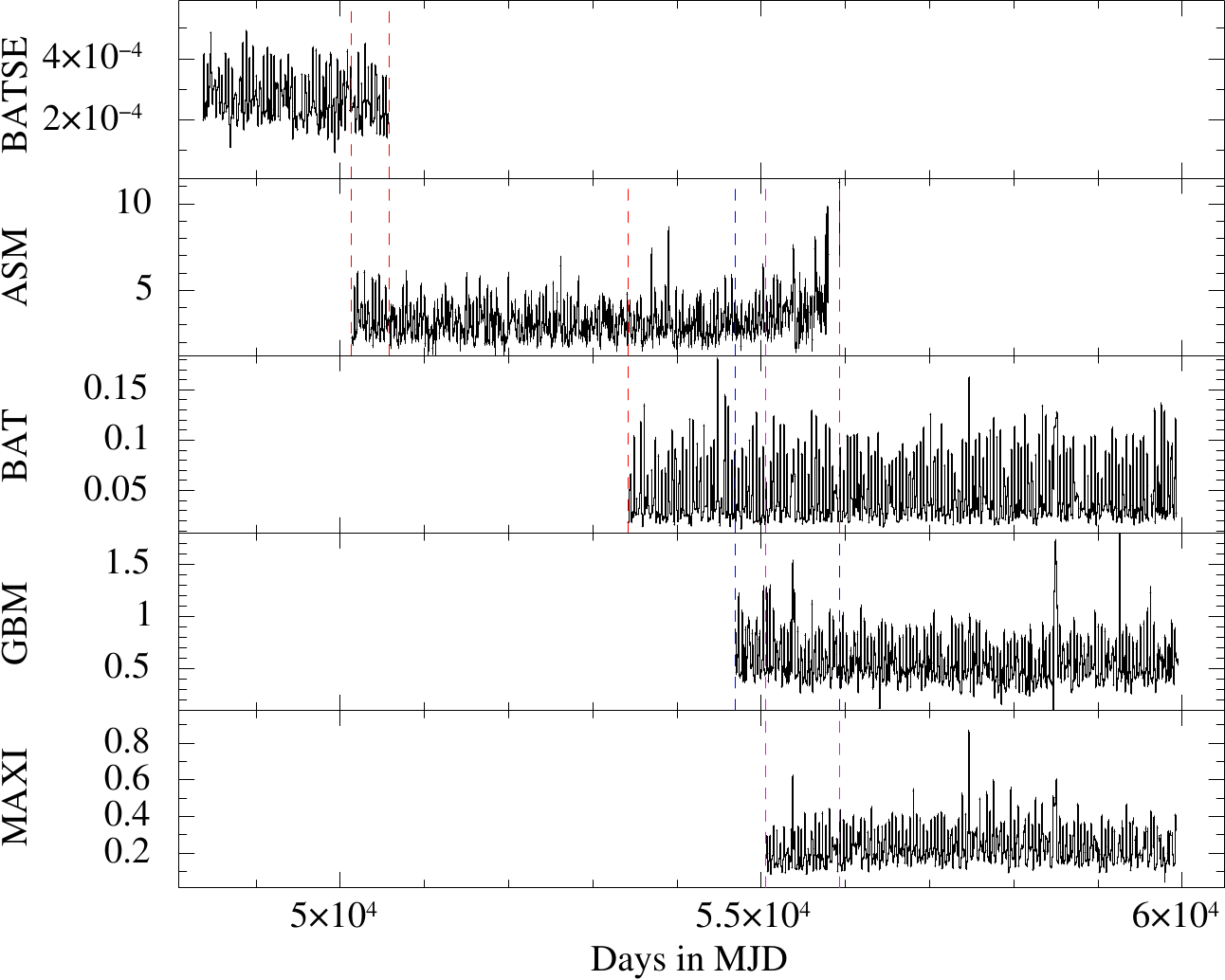}
    \caption{The long term lightcurves and pulsed flux histories from different All-sky monitors plotted with a bin size of 10 d. The overlapping duration for BATSE--ASM, BAT--ASM, BAT--GBM and BAT--\textit{MAXI} are represented with vertical dashed lines. The simultaneous data allowed a check for the energy dependence of flares, and we found a clear hard lag (Table~\ref{tab:overlap_lc_2}).}
    \label{fig:ltlc}
\end{figure}

However, the five long-term lightcurves are from different energy ranges, and the periodic pre-periastron flares of GX 301--2 are known to exhibit a hard X-ray lag of about a day \citep{liu_gx301_hardlag}. Therefore, we checked the simultaneity of the flare peaks in the long-term lightcurves before proceeding with the $\dot P$\textsubscript{orb} estimation. The long-term lightcurves and pulsed histories have overlapping data duration (See Table~\ref{tab:overlap_lc_2} and the vertical dashed lines in Fig.~\ref{fig:ltlc}). We checked the flares in BATSE (20--50 keV), ASM (1.5--12 keV), BAT (15--50 keV), GBM (12--50 keV) and \textit{MAXI} (4--10, 10--20 keV). The long-term lightcurve from \textit{Swift}/BAT has considerable overlapping data duration with \textit{RXTE}/ASM, \textit{Fermi}/GBM, and \textit{MAXI} lightcurves, and \textit{CGRO}/BATSE has overlap with \textit{RXTE}/ASM to perform this study. We estimated the difference in flare times ($\Delta T$\textsubscript{flare}) between lightcurves in the overlapping durations using the technique described in Appendix~\ref{appendix:flareEdep}. Except for BATSE and GBM, we found a very clear hard X-ray lag of $\sim$ 0.9 d (Table~\ref{tab:overlap_lc_2}). As the BATSE and GBM pulsed flux histories are generated by integrating the pulsed flux over one day and two days, respectively, which is of the same order as the flare duration, it could impact the accurate construction of the flare shape and, subsequently, our estimation of the flare peak. This inadequacy of the data most likely causes the contrasting results from BATSE and GBM. We derived error scaling factors for the flare times $T_n$ for BATSE (4.3), GBM (5.8) and \textit{MAXI} 10--20 keV (4), and a time shift for ASM ($+0.96$ d) so that the energy dependence of flare arrival times are eliminated and all the flare times are consistent with BAT.

\begin{table*}
    \caption{Peak flare times from overlapping duration of the long-term lightcurves in different energy bands.}
    \centering
    \begin{tabular}{cccccc}
        \hline
        \hline
         Observatory/Instrument &Energy range (keV) &LC duration (MJD) &No. of orbits during overlap &$\Delta T_\textrm{flare}$ (d)\\
         \hline
        \textit{CGRO}/BATSE &20-50 keV &48370-50579 &\multicolumn{2}{c}{Reference LC}\\
        \textit{RXTE}/ASM &1.5--12 keV &50133-55927 &10 &$-0.57\pm0.09$\\
        \\
         \textit{Swift}/BAT &15--50 keV &53416-59927 &\multicolumn{2}{c}{Reference LC}\\
         \textit{RXTE}/ASM &1.5--12 keV &50133-55927  &60 &$-0.96\pm0.06$\\
         \textit{Fermi}/GBM &12--50 keV  &54691-59947 &126 &$-0.23\pm0.04$\\
         \textit{MAXI} &2--20 keV  &55053-59927 &116 &$-0.36\pm0.02$\\
         &4--10 keV &" &" &$-0.88\pm0.03$\\
         &10--20 keV &" &" &$-0.08\pm0.02$\\
         \hline
    \end{tabular}
    \label{tab:overlap_lc_2}
\end{table*}

\begin{figure}
    \centering
    \includegraphics[width=\columnwidth]{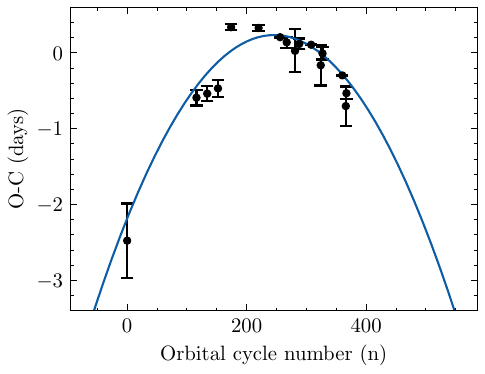}
    \includegraphics[width=\columnwidth]{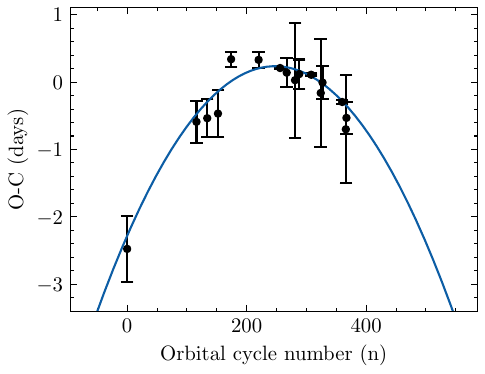}
    \caption{O--C curve from the pre-periastron flare peak times derived from BATSE, ASM, BAT, GBM and \textit{MAXI}. The first data point is taken from \protect\cite{sato1986-orb}. Error in GBM data points were scaled by a factor of 8.3, and ASM data points were shifted by +0.7 days to account for the energy dependence of flare times (See text). Top figure shows the quadratic fit indicating $\dot P_\textrm{orb} = -(1.93\pm0.11)\times10^{-6}$ s \textsuperscript{-1}. However, the weighted variance (\texttt{wvar}) of the fit was poor at 145 for 14 ($16-3+1$) d.o.f, which impacts the parameter error estimation. The large variance was contributed by the low error bar of the data points ($\sum_{i=1}^{15}(\frac{d_i-m_i}{e_i})^2\sim145)$. To make the wvar$\approx$d.o.f, we scaled up each error $e_i$ with a scaling factor of $\sqrt{{145}/{15}}\sim3$. This reduced the \texttt{wvar} to $\sim16$ (14 d.o.f). The bottom figure shows the best-fit quadratic model on the error re-scaled data. Best fit $\dot P_\textrm{orb}$ is $-(1.98\pm0.28)\times10^{-6}$ s s$^{-1}$. The quoted errors on all the parameters are their 2.7$\sigma$ confidence ranges.}
    \label{fig:quadrfit}
\end{figure}

The time stamps of the pre-periastron flare $T_n$s derived from the long-term lightcurves were corrected for the energy dependence mentioned before and the energy independent flare times were used for further analysis. In addition to these data points, we used the flare time from \cite{sato1986-orb}, which was derived from \textit{Ariel-V} (2--15 keV), \textit{SAS--3} (8--18 keV), \textit{Hakucho} (9--22 keV) and \textit{HEAO--1} (15--175 keV). The $T_n$ vs $n$ was then fitted with a linear function in $n$ (mimicking $T_0 + nP_\textrm{orb}$), and the residuals to the linear fit ($\delta T_n$ or O--C vs $n$) were plotted. A clear negative parabolic trend was visible in the residuals, indicating the orbital decay (Fig.~\ref{fig:quadrfit}). Fitting a function of the form $T_0 + nP_\mathrm{orb}+ 0.5n^2P_\textrm{orb}\dot P_\textrm{orb}$ gave the best fit $\dot P_\textrm{orb}$ as $-(1.93\pm0.11)\times10^{-6}$ s s$^{-1}$ (Fig.~\ref{fig:quadrfit} top). However, the fit-statistic was large, and we, therefore, scaled the errors in flare times by a factor of 3. This is justified because, along with the regular pre-periastron  flares,  GX 301–2 is also known to exhibit short-term variability, which could contribute to additional systematic error in the determination of pr-periastron flare times. Scaling of errors improved the fit statistic and the subsequently obtained best fit $\dot P_\textrm{orb}$ is $-(1.98\pm0.28)\times10^{-6}$ s s$^{-1}$ (Fig.~\ref{fig:quadrfit} bottom).

\section{Discussions}

\subsection{Estimation of decay in orbital period}

The rapid orbital decay rate of GX 301--2 was estimated by \cite{doroshenkov2010-orb} from multiple \textit{INTEGRAL} pointed observations by timing the X-ray pulses. A constant $\dot P$\textsubscript{spin} of the pulsar was assumed in the calculation. However, GX 301--2 exhibits an intensity variation by a factor of 15 within the orbit (evident from folded \textit{Swift}/BAT orbital intensity profile in Fig.~\ref{fig:ltlc_orbit_profiles}) and even a factor of 3 during the out-of-flare states \citep{gx301_nustar_felix}. The torque state of the X-ray pulsar is also known to be dependent on its luminosity \cite{Pravdo_2001}. These factors adversely impact the assumption of a constant $\dot P$\textsubscript{spin} and subsequently the estimation of $\dot P$\textsubscript{orb} \citep{2020monkkoen}. 

Our estimate of the orbital period decay from an independent method using the flare timing signatures in long-term X-ray lightcurves is not affected by the uncertainty of $\dot P$\textsubscript{spin}. Assuming the individual flare peaks are accurate to $\delta t\sim0.0625$ d, for a time interval of $\Delta t\sim10^4$ d, an orbital evolution timescale $|t_p|=|P_\mathrm{orb}/\dot P_\mathrm{orb}|\sim10^5$ yr could be estimated to a precision of $t_p\delta t/\Delta t^2 \times 100\sim$ 2\% \citep{peter_eggleton}. However, this technique of $\dot P$\textsubscript{orb} estimation will depend on the shape of the orbital intensity profile, which has the major contribution from the pre-periastron flare and will be the main contributor to the uncertainty of this technique. Even though not entirely understood, the orbital profile of GX 301--2 is generally explained on the basis of two common models by (i) \cite{Pravdo_2001} based on an equatorial circumstellar disc of gas around the companion star and (ii) \cite{gx301_gas_stream_haberl,LC_stream_model} based on a dense stream of matter from the companion following the pulsar. Changes in the properties of the circumstellar disk or the accretion stream could therefore result in variations in the shape of the orbital intensity profile. Our analysis is the most accurate if the orbital intensity profile stays the same throughout the long-term data used for the analysis.

Our analysis also suggests the presence of a rapid orbital decay. The estimate of orbital period decay is $\dot P_\textrm{orb}=-(1.98\pm0.28)\times10^{-6}$ s s$^{-1}$ corresponding to an orbital evolution time scale of $|P_\textrm{orb}/\dot P_\textrm{orb}| \approx 0.6\times10^5$ yr. Our estimate of $\dot P$\textsubscript{orb} is different from the value reported by \cite{doroshenkov2010-orb}, which is $-(3.7\pm0.5)\times10^{-6}$ s s$^{-1}$, by a factor of $\sim2$.

\subsection{Possible reasons for the rapid orbital decay}\label{possibilities}

The observed orbital evolution time scale of $|P_\textrm{orb}/\dot P_\textrm{orb}|$ $\sim10^{5}$ years in GX 301--2  is an order of magnitude shorter than the mass loss time scale of the companion of $|{M_c}/{\dot M_c}|\sim10^{6}$ years. Until now, this is the fastest orbital decay ever observed in an HMXB (See Table~\ref{tab:hmxb_orbevolve}). Even though there was a recent report by \cite{cenx3_orbitalP_astrosat} of a much larger orbital decay rate of $|\dot P_\textrm{orb}/P_\textrm{orb}|\sim10^{-4}$ yr$^{-1}$ in the HMXB Cen X--3, which contradicts previous measurements (Table~\ref{tab:hmxb_orbevolve}), it should be noted that for the pulse time-of-arrival (TOA) analysis, the authors utilized data from only a portion (half) of one orbit, and any intrinsic variations in the pulsar spin rate may have contributed to this disparate result. Disregarding this report, GX 301--2 has exhibited the fastest observed orbital decay among HMXBs, and we are examining potential causes for the observed orbital decay.

The orbital evolution of a binary star system can be described by the changes in its orbital angular momentum  and mass transfer \citep{tauris_vandenheuvel_2006,Bachetti2022-M82X2-orbdecay} as follows 
\begin{equation}
    \frac{2}{3}\frac{\dot P_\textrm{orb}}{P_\textrm{orb}} = 2\frac{\dot J_\textrm{orb}}{J_\textrm{orb}} - 2\frac{\dot M_c}{M_c} - 2\frac{\dot M_x}{M_x} + \frac{\dot M_c + \dot M_x}{M_c + M_x} - 2e\dot e\label{eq1}
\end{equation} 

In equation~\ref{eq1}, the binary orbital period $P_\mathrm{orb}$ and its rate of change $\dot P_\mathrm{orb}$ are expressed in terms of the evolution of other binary parameters. $J_\textrm{orb}$ and $\dot J_\textrm{orb}$ are the orbital angular momentum of the binary and its rate of change, respectively, $M_c$ and $\dot M_c$ are the companion mass and its rate of change, respectively, and $M_x$ and $\dot M_x$ are the NS mass and its rate of change, respectively. 

Some of these parameters are known for GX 301--2 (Table~\ref{tab:knownpars}). The observed orbital decay ($\dot P$\textsubscript{orb}$<0$) in GX 301--2 could be investigated through equation~\ref{eq1}, which implies that $\dot J$\textsubscript{orb}$<0$, $\dot e>0$ and certain combinations of $\dot M_c$, $\dot M_x$, $M_x$ and  $M_c$ has the potential to cause $\dot P$\textsubscript{orb}$<0$. Furthermore, some of these parameters may exert a greater influence on $\dot P$\textsubscript{orb} compared to the others. A case in point is, although  equation~\ref{eq1} suggests that $\dot e<0$ can lead to the expansion of the orbit, in HMXBs the opposite is observed. The reason could be that $\dot e<0$ in HMXBs arises from tidal interactions, which also results in $\dot J$\textsubscript{orb}$<0$ and the $\dot J$ term dominates over the $\dot e$ term, causing the orbit to decay instead of expanding.

Our aim is to evaluate three feasible factors that could produce the observed orbital decay in GX 301--2, which are mass transfer from the companion to the NS, mass loss from the binary, and tidal interaction between NS and the companion. The conservation of $J_\textrm{orb}$ characterizes the former mechanism, in which the decay of the orbit is driven by mass redistribution. On the other hand, the latter two mechanisms are characterized by loss of $J_\textrm{orb}$, leading to the decay of the orbit. Recent simulations of GX 301--2 by \cite{Bunzel2022-simulation} do not predict this rapid orbital decay before the Common Envelope phase, but not all of the aforementioned mechanisms were included in their simulations. Although the loss of $J$\textsubscript{orb} is also possible due to gravitational wave radiation and magnetic braking, they are only dominant in orbits that are sufficiently compact, as stated by \cite{interactingbinaries_vandenheuvel}, and hence we do not discuss it further.

\begin{table}
    \centering
    \caption{Some reported estimates of GX 301--2 parameters. We used these values to assess various possibilities of the observed orbital decay in Section~\ref{possibilities}.}
    \begin{tabular}{l|cl}
        \hline
        \hline
         Parameter &Value &Reference\\
         \hline
         $P_\textrm{orb}$ &41.5 d & \cite{sato1986-orb}\\
         $|\dot P_\textrm{orb}/P_\textrm{orb}|$ &$5.52\times10^{-13}$ s$^{-1}$ &This work\\
         &$1.74\times10^{-5}$ yr$^{-1}$ &\\
         $M_x$ &1.4 M$_\odot$ &Canonical\\
         $M_c$ & 50 M$_\odot$ &\cite{kaper_massloss_estimate_1995}\\
         $R_c$ &87 R$_\odot$ &\cite{kaper_massloss_estimate_1995}\\
         $i$ &$\le64^\circ$ &\cite{kaper_massloss_estimate_1995}\\
         $a_x\textrm{sin}\ i$ &$159\pm1.5$ R$_\odot$ &\cite{sato1986-orb}\\
         $a_x$ &177 R$_\odot$\\
         $e$ &0.47 &\cite{sato1986-orb}\\
         $\dot M_c$ &$-(3\ \textrm{to}\ 10)\times10^{-6}$ M$_\odot$ yr$^{-1}$ &\cite{parkes_gx301_windvelo_windrate_1980};\\
         &&\cite{kaper_massloss_estimate_1995}\\
         $v$\textsubscript{wind} &400 km s$^{-1}$ &\cite{parkes_gx301_windvelo_windrate_1980}\\
         $v_e$sin $i$& 55 km s$^{-1}$ &\cite{BPCru_Clarke}\\
         $v_e$ &61 km s$^{-1}$\\
         $P_c^\dagger$ &72 d\\
         \hline
    \end{tabular}
    \begin{tablenotes}
        \item $^\dagger$ $P_c=2\pi R_c/v_e$.
    \end{tablenotes}
    \label{tab:knownpars}
\end{table}

\subsubsection{Conservative mass transfer}

The simplest case is the conservative mass transfer from companion to the NS, where the orbital angular momentum is conserved ($\dot J_\textrm{orb}=0$), and eccentricity stays constant ($\dot e=0$). In the scenario of conservative mass transfer, the entire mass lost by the companion is accreted by the neutron star ($-\dot M_c = M_x$), and there is no significant alteration of the orbital angular momentum ($\dot J_\textrm{orb} = 0$).

Substituting the values from Table~\ref{tab:knownpars} in equation~\ref{eq:en}, the required mass transfer rate (accretion rate) to the NS for attaining the observed orbital decay rate is $\dot M_x$ $\sim8\times10^{-6}\ \mathrm{M}_\odot$ yr$^{-1}$. This is roughly the mass loss rate from the companion (Table~\ref{tab:knownpars}). However, the Eddington accretion limit for spherical accretion of Hydrogen-rich matter to a canonical 1.4 M$_\odot$ 10 km radius NS is about $10^{-8}$ M$_\odot$ yr$^{-1}$ \citep{interactingbinaries_vandenheuvel}, implying only a maximum of $\sim1\%$ of the mass lost by Wray 15-977 could be accreted by the NS even if it is accreting at the Eddington limit. Therefore, conservative mass transfer can't be the primary mechanism driving the observed $\dot P$\textsubscript{orb} in GX 301--2.

\subsubsection{Mass loss from the binary}
 
The efficiency of wind accretion in GX 301--2 could be calculated using the equations $e_\textrm{wind}=\pi r_\textrm{acc}^2/4\pi a_x^2$ and $r_\textrm{acc}=GM_x/v_w^2$. Here $e_\textrm{wind}$ is the efficiency of wind accretion, $v_w$ is the velocity of stellar wind from the companion, and accretion radius $r_\textrm{acc}$ is the distance from the NS at which the stellar wind is gravitationally captured. Substituting values for GX 301--2 from Table~\ref{tab:knownpars} gives the efficiency of wind accretion $e_\textrm{wind}\sim3\times10^{-5}$. The unaccreted matter will likely be lost from the binary and contribute to $\dot J_\textrm{orb}$. A complete consideration of mass loss from the binary makes the estimation of binary evolution a three-body problem ($M_x$, $M_c$ and the lost mass $\delta M$), rendering a general solution difficult. Therefore, certain physically motivated scenarios for loss of mass from the binary (mass loss modes) causing $\dot J$\textsubscript{orb} viz., Jeans' mode, Isotropic re-emission mode and Intermediate mode (See \citealt{huang_1963_masslossmodes} and \citealt{interactingbinaries_vandenheuvel}) are usually explored. If the mass loss from the binary is the most dominant factor contributing to $\dot J$\textsubscript{orb}, assuming a mass loss to proceed in any of these three mentioned modes, $\dot J$ can be expressed as (equation~16.18 in \citealt{tauris_vandenheuvel_2006}): 

\begin{equation}
    \frac{\dot J}{J}=\frac{\alpha + \beta q^2 + \delta\gamma(1+q^2)}{1+q}\frac{\dot M_c}{M_c}\label{eq2}
\end{equation}
\begin{equation}
    \dot M_x = -(1-\alpha-\beta-\delta)\dot M_c\label{eq3}
\end{equation}

where $\alpha,\beta$ and $\delta$ denote the fractions of mass lost from the companion by (i) direct isotropic wind without gravitationally interacting with the NS (Jean's mode), (ii) isotropic ejection after being captured by the NS gravitational field (Isotropic re-emission), and (iii) lost mass overcoming the individual gravitational attractions of companion and NS, and escape through the lagrangian points $L_2$ or $L_3$ to form an extended circumbinary ring revolving around the common mass ($M_c+M_x$) of binary at a radius of $\gamma^2a_x$ (Intermediate mode), respectively. $q=M_c/M_x$ is the mass ratio and $\epsilon = 1-\alpha-\beta-\delta$ denotes the fraction of mass accreted. 

Individual contributions to orbital evolution due to these three different modes of mass loss could be explored by assigning values for $\alpha,\beta$, and $\delta$ and using the equations \ref{eq1},\ref{eq2} and \ref{eq3}, assuming $\dot e=0$. 

A direct isotropic wind loss from the companion could be defined by ($\alpha=1,\beta=\delta=0$). If the lost mass has an outward velocity greater than the escape velocity, it emulates an instant reduction of the total mass in the binary and hence the gravitational attraction between two stellar components. This leads to expansion of the orbit ($\dot P_\textrm{orb}>0$) instead of the observed orbital decay. Simulations of the wind loss from Wray 15-977 indeed show this physical scenario causing expansion of the orbit in GX 301--2 (Fig.~5 of \citealt{Bunzel2022-simulation}).

Isotropic re-emission from the vicinity of the NS could be defined by ($\alpha=0,\beta=1,\delta=0$). In this case, the mass lost by the stellar wind from Wray 15-977 is first conservatively captured by the gravitational pull of NS and then re-ejected isotropically from the vicinity of NS. The re-emission of matter could occur due to radiation/magnetically driven wind from the neutron star as pointed out by \cite{doroshenkov2010-orb}. This scenario can lead to orbital decay. Substituting known values from Table~\ref{tab:knownpars} demonstrates that the observed orbital decay can occur for $\dot M_c\sim9\times10^{-6}$ M$_\odot$ yr$^{-1}$. Despite the scenario being considered, it cannot fully account for the observed orbital evolution in GX 301--2 because of the companion's inability to undergo a conservative mass transfer to the NS vicinity due to the poor wind capture efficiency ($e_\mathrm{wind}<<1$). 

Anisotropic mass loss from the companion through $L_2$ or $L_3$ resulting in the formation of an extended toroidal ring around the common mass ($M_c+M_x$) at a distance $\gamma^2a_x$ from the centre of mass could be defined by ($\alpha=\beta=0,\delta=1$). For $\gamma\gtrapprox1$, $\dot M_c\le9\times10^{-6}$ M$_\odot$ yr$^{-1}$ have the potential to produce the observed orbital decay in GX 301--2.

Although each mass loss mode alone could not be responsible for the observed orbital decay, it is possible that the actual mass ejection mode could be a composite of these idealized modes, and thereby produce the observed orbital decay.

\subsubsection{Tidal interaction}

Apart from mass-loss from the binary, another dominant mechanism that can contribute to $\dot J$\textsubscript{orb} is tidal interaction between the NS and the rotating deformable companion in an eccentric binary (\citealt{darwin_tide}; \citealt{lecar_tidal_circularize_herx1_cenx3}). The compact object raises a tide on the surface of the companion. The tide facilitates angular momentum exchange between the rotating companion and the binary orbit and the dissipation of rotational and orbital energies. This results in synchronising the rotation of the companion and binary orbit (tidal synchronization) and circularizing the binary orbit (tidal circularization). If the companion rotation frequency ($\Omega_c$) is less than the binary orbital frequency ($\Omega$\textsubscript{orb}), the retarding force of tide at the periastron is expected to circularize the orbit and cause orbital decay in the process\footnote{One could grasp in a general sense the tide induced orbital decay, based on the principle of Hohmann orbit for satellite transfer \citep{hohmann1960attainability}, even though both phenomena are unrelated.}. The spin angular momentum of the companion will increase at the expense of orbital angular momentum in this scenario. 

A general form of tidal evolution in an HMXB is rather complex, which includes invoking dynamical tides that cause oscillating tidal response from the companion \citep{witte_tide}. However, a fairly simple approximation is the weak friction model of the tide which does not include the non-linear tidal dissipation processes (Refer \citealt{tidal_hut1981}). Our objective is to comprehend the swift orbital evolution witnessed in GX 301--2 concerning tidal dissipation through the weak friction model. Calculations based on \cite{lecar_tidal_circularize_herx1_cenx3} and \cite{tidal_hut1981} under the assumption of weak friction model shows that tidal dissipation in the outer convective envelope of Wray 15-977 having a characteristic $\lambda\eta v_\textrm{conv} = 2\times10^{-4}$ km s$^{-1}$ ($\lambda$ is the fractional depth of the convective layer of the companion, $\eta$ is the fractional mass of the convective layer, and $v$\textsubscript{conv} is the convective velocity) can cause the observed orbital decay in GX 301--2 (See Appendix~\ref{Appx:tidalcirctimescale} for detailed calculation). Considering the significant mass loss rate of the companion which can cause expansion of the binary orbit, the calculated convective envelope parameters would be a lower limit if tidal dissipation is the lone factor driving orbital decay in GX 301--2.

A complete consideration of the effect of tidal interaction invoking the dynamical tides to estimate the tidal parameters required to produce the observed rapid orbital evolution of GX 301--2 is beyond the scope of this work. However, we refer to the work \cite{donglai_psrj0045} which discusses the orbital decay of the young eccentric binary radio pulsar PSR J0045-7319 having similar binary parameters as GX 301--2 ($P$\textsubscript{orb}$\sim52$ d, $e\sim0.8$, B-type $M_c\sim9$M$_\odot$, $a_x\sim12$R$_\odot$) and exhibits a rapid orbital decay of $|P_\textrm{orb}/\dot P_\textrm{orb}|\sim5\times10^5$ yr. \cite{donglai_psrj0045} had shown that tidal interaction between the pulsar and a retrograde spinning companion may cause such a rapid orbital decay by invoking dynamical tides.

In binary systems such as GX 301--2, where there exists a significant difference in the mass of the components, with the mass ratio $M_c/M_x\sim35$, it is possible for the system to undergo a Common Envelope (CE) phase during the later stages of evolution, due to either tidally induced orbital decay or significant Roche lobe overflow. Tidal interactions proceed towards synchronizing the slow rotation of the companion star with the fast binary orbit. However, in situations where the companion star is significantly more massive than the neutron star, the latter finds it difficult to spin up the former. An intriguing outcome occurs when $J_\textrm{orb}\lesssim3J_\textrm{c}$ (equations 102 and 99 in \citealt{interactingbinaries_vandenheuvel}), where the binary orbit continues to shrink, gradually achieving synchronization with the slowly spinning, massive companion, culminating in `tidal catastrophe' where the neutron star spirals towards the core of the companion and merges.

Assuming an optimal scenario in which the binary orbit synchronises with the companion by the time of circularization, i.e., $\Omega_c=\Omega_\textrm{orb}=\Omega$. The relation $3J_\textrm{c}/J_\textrm{orb}>1$ can be simplified to $3I_\textrm{c}/I_\textrm{orb}>1$ (See \citealt{lecar_tidal_circularize_herx1_cenx3}), where $I$\textsubscript{c} and $I$\textsubscript{orb} represents the moment of inertia of the companion and binary orbit, respectively, at the later circularized phase. Since the orbital separation is expected to shrink by this time, $I_\textrm{orb}\lesssim M_xa_x^2\lesssim 4.4\times10^{4}$ M$_\odot$R$_\odot^2$. Meanwhile, the companion star is expected to evolve, resulting in an increase in its radius and a decrease in mass due to stellar wind. Assuming $I_\textrm{c}\approx M_cR_c^2\approx 40\times10^{4}$ M$_\odot$R$_\odot^2$. The ratio $3I_\textrm{c}/I_\textrm{orb}$ is $\gtrsim30$, indicating an unstable orbit post orbit circularization and the possibility of tidal catastrophe.

In the Roche lobe overflow phase, if the NS cannot accept the Roche lobe overflown matter from the companion beyond the Eddington accretion rate, it forms a Common Envelope (CE) surrounding both stars. This CE phase can also result in the spiral in of NS due to frictional drag in the companion's stellar envelope, as proposed by \cite{Bunzel2022-simulation} for GX 301--2. The aftereffect of the CE phase could be the ejection of the common envelope and subsequent formation of a binary comprising the already existing neutron star and the companion's He-rich core. However, if the orbital energy lost during spiralling-in is not efficiently converted into mechanical energy and transferred to the envelope for CE ejection, it may instead result in the NS merging with the core of the companion. The resulting unique object has a NS core surrounded by H/He envelope \citep{interactingbinaries_vandenheuvel} and is called Thorne-\.Zytkow Object (TZO) \citep{thorne1977stars}. The same may happen with the Tidal catastrophe as well. GX 301--2 is thus a prospective future TZO candidate.

\begin{table}
    \centering
    \caption{Previous reports of the orbital decay reported for HMXBs in the order of increasing $|\dot P_\textrm{orb}/P_\textrm{orb}|$. The evolution time scale is of the order of the inverse of the second column. The shortest evolution timescale corresponds to GX 301--2 ($\sim10^5$ yr), and the longest evolution timescale corresponds to OAO 1657--415 ($\sim10^{7}$) yr.}
    \begin{tabular}{lll}
        \hline
        \hline
         Source &$\dot P_\textrm{orb}/P_\textrm{orb}$ (in $10^{-6}$ yr\textsuperscript{-1}) &Reference\\
         \hline
         OAO 1657--415 &$-0.0974\pm0.0078$ &\cite{oao1657_orbPdot_jenke}\\
         4U 1700--37 &$-0.47\pm0.19$ &\cite{4u1700_orbPdot_islam2016}\\
         4U 1538--52 &$-0.95\pm0.37$ &\cite{4u1538_orbPdot}\\
         Cyg X--3 &$-1.05\pm0.04$ &\cite{singh_cygx3}\\
         LMC X--4 &$-0.989\pm0.005$ &\cite{lmcx4_orbPdot_Naik_2004}\\
         Cen X--3 &$-1.799\pm0.002$ &\cite{cenx3_orbPdot_raichur_rxte}\\
         SMC X--1 &$-3.414\pm0.003$ &\cite{cenx3_orbPdot_raichur_rxte}\\
         GX 301--2 &$-32.5\pm4.4$ &\cite{doroshenkov2010-orb}\\
         &$-17.4\pm2.5$&This work\\
         \hline
    \end{tabular}
    \label{tab:hmxb_orbevolve}
\end{table}

\section{Conclusions}

In this study, we utilized the recurring pre-periastron flares observed in the long-term X-ray lightcurves of GX 301--2 to measure its orbital period evolution. Our analysis yielded a measured orbital decay timescale of $|\dot P_\textrm{orb}/P_\textrm{orb}|$ $\sim2\times 10^{-5}$ yr$^{-1}$, which is currently the shortest known evolution timescale for a high-mass X-ray binary (HMXB). Previous estimates of this decay timescale were based on pulse time-of-arrival (TOA) analysis, which is influenced by the large orbital intensity variations and spin-up/down fluctuations of the pulsar. Our analysis of the long-term lightcurves, however, relies on the recurring orbital intensity profile, which is independent of the pulse TOA methods. There is a difference of about a factor of two between our estimate and the previous estimate based on pulse TOA analysis. Our estimate is limited by the repeatability of pre-periastron flares and is dependent on the stability and recurrence of the process causing these flares, which is still uncertain. We argue that a combination of distinct mechanisms, such as unique mass loss pathways and/or tidal interaction could be driving this rapid orbital decay.

\section*{Acknowledgements}

 We acknowledge the use of public data from the Swift data archive. This research has made use of MAXI data provided by RIKEN, JAXA and the MAXI team. This research has made use of data and/or software provided by the High Energy Astrophysics Science Archive Research Center (HEASARC), which is a service of the Astrophysics Science Division at NASA/GSFC. We acknowledge the use of quick-look results provided by the ASM/RXTE team. We thank the anonymous referee for helpful suggestions.

%%%%%%%%%%%%%%%%%%%%%%%%%%%%%%%%%%%%%%%%%%%%%%%%%%
\section*{Data Availability}

All the data underlying this research article are publicly available for download from the respective mission web pages.



\bibliographystyle{mnras}
\bibliography{example} % if your bibtex file is called example.bib

\begin{thebibliography}{}
\makeatletter
\relax
\def\mn@urlcharsother{\let\do\@makeother \do\$\do\&\do\#\do\^\do\_\do\%\do\~}
\def\mn@doi{\begingroup\mn@urlcharsother \@ifnextchar [ {\mn@doi@}
  {\mn@doi@[]}}
\def\mn@doi@[#1]#2{\def\@tempa{#1}\ifx\@tempa\@empty \href
  {http://dx.doi.org/#2} {doi:#2}\else \href {http://dx.doi.org/#2} {#1}\fi
  \endgroup}
\def\mn@eprint#1#2{\mn@eprint@#1:#2::\@nil}
\def\mn@eprint@arXiv#1{\href {http://arxiv.org/abs/#1} {{\tt arXiv:#1}}}
\def\mn@eprint@dblp#1{\href {http://dblp.uni-trier.de/rec/bibtex/#1.xml}
  {dblp:#1}}
\def\mn@eprint@#1:#2:#3:#4\@nil{\def\@tempa {#1}\def\@tempb {#2}\def\@tempc
  {#3}\ifx \@tempc \@empty \let \@tempc \@tempb \let \@tempb \@tempa \fi \ifx
  \@tempb \@empty \def\@tempb {arXiv}\fi \@ifundefined
  {mn@eprint@\@tempb}{\@tempb:\@tempc}{\expandafter \expandafter \csname
  mn@eprint@\@tempb\endcsname \expandafter{\@tempc}}}

\bibitem[\protect\citeauthoryear{Bachetti et~al.,}{Bachetti
  et~al.}{2022}]{Bachetti2022-M82X2-orbdecay}
Bachetti M.,  et~al., 2022, \mn@doi [The Astrophysical Journal]
  {10.3847/1538-4357/ac8d67}, 937, 125

\bibitem[\protect\citeauthoryear{Barthelmy et~al.,}{Barthelmy
  et~al.}{2005}]{bat_2005}
Barthelmy S.~D.,  et~al., 2005, Space Science Reviews, 120, 143

\bibitem[\protect\citeauthoryear{{Boldin}, {Tsygankov}  \&
  {Lutovinov}}{{Boldin} et~al.}{2013}]{boldin_2013_efs_error}
{Boldin} P.~A.,  {Tsygankov} S.~S.,   {Lutovinov} A.~A.,  2013, \mn@doi
  [Astronomy Letters] {10.1134/S1063773713060029}, \href
  {https://ui.adsabs.harvard.edu/abs/2013AstL...39..375B} {39, 375}

\bibitem[\protect\citeauthoryear{Bunzel, Garc{\'\i}a, Combi  \& Chaty}{Bunzel
  et~al.}{2023}]{Bunzel2022-simulation}
Bunzel A.~S.,  Garc{\'\i}a F.,  Combi J.~A.,   Chaty S.,  2023, Astronomy \&
  Astrophysics, 670, A80

\bibitem[\protect\citeauthoryear{{Clark}, {Najarro}, {Negueruela}, {Ritchie},
  {Urbaneja}  \& {Howarth}}{{Clark} et~al.}{2012}]{BPCru_Clarke}
{Clark} J.~S.,  {Najarro} F.,  {Negueruela} I.,  {Ritchie} B.~W.,  {Urbaneja}
  M.~A.,   {Howarth} I.~D.,  2012, \mn@doi [\aap]
  {10.1051/0004-6361/201117472}, \href
  {https://ui.adsabs.harvard.edu/abs/2012A&A...541A.145C} {541, A145}

\bibitem[\protect\citeauthoryear{{Darwin}}{{Darwin}}{1879}]{darwin_tide}
{Darwin} G.~H.,  1879, The Observatory, \href
  {https://ui.adsabs.harvard.edu/abs/1879Obs.....3...79D} {3, 79}

\bibitem[\protect\citeauthoryear{{Doroshenko}, {Santangelo}, {Suleimanov},
  {Kreykenbohm}, {Staubert}, {Ferrigno}  \& {Klochkov}}{{Doroshenko}
  et~al.}{2010}]{doroshenkov2010-orb}
{Doroshenko} V.,  {Santangelo} A.,  {Suleimanov} V.,  {Kreykenbohm} I.,
  {Staubert} R.,  {Ferrigno} C.,   {Klochkov} D.,  2010, \mn@doi [\aap]
  {10.1051/0004-6361/200912951}, \href
  {https://ui.adsabs.harvard.edu/abs/2010A&A...515A..10D} {515, A10}

\bibitem[\protect\citeauthoryear{{Eggleton}}{{Eggleton}}{2006}]{peter_eggleton}
{Eggleton} P.,  2006, {Evolutionary Processes in Binary and Multiple Stars}

\bibitem[\protect\citeauthoryear{{F{\"u}rst} et~al.,}{{F{\"u}rst}
  et~al.}{2018}]{gx301_nustar_felix}
{F{\"u}rst} F.,  et~al., 2018, \mn@doi [\aap] {10.1051/0004-6361/201732132},
  \href {https://ui.adsabs.harvard.edu/abs/2018A&A...620A.153F} {620, A153}

\bibitem[\protect\citeauthoryear{Gehrels et~al.,}{Gehrels
  et~al.}{2004}]{Swift_Gehrels_2004}
Gehrels N.,  et~al., 2004, \mn@doi [The Astrophysical Journal]
  {10.1086/422091}, 611, 1005

\bibitem[\protect\citeauthoryear{{Haberl}}{{Haberl}}{1991}]{gx301_gas_stream_haberl}
{Haberl} F.,  1991, \mn@doi [\apj] {10.1086/170273}, \href
  {https://ui.adsabs.harvard.edu/abs/1991ApJ...376..245H} {376, 245}

\bibitem[\protect\citeauthoryear{Hemphill et~al.,}{Hemphill
  et~al.}{2019}]{4u1538_orbPdot}
Hemphill P.~B.,  et~al., 2019, \mn@doi [The Astrophysical Journal]
  {10.3847/1538-4357/ab03d3}, 873, 62

\bibitem[\protect\citeauthoryear{Hohmann}{Hohmann}{1960}]{hohmann1960attainability}
Hohmann W.,  1960, The attainability of heavenly bodies.
No.~44, National Aeronautics and Space Administration

\bibitem[\protect\citeauthoryear{{Huang}}{{Huang}}{1963}]{huang_1963_masslossmodes}
{Huang} S.-S.,  1963, \mn@doi [\apj] {10.1086/147659}, \href
  {https://ui.adsabs.harvard.edu/abs/1963ApJ...138..471H} {138, 471}

\bibitem[\protect\citeauthoryear{{Hut}}{{Hut}}{1981}]{tidal_hut1981}
{Hut} P.,  1981, \aap, \href
  {https://ui.adsabs.harvard.edu/abs/1981A&A....99..126H} {99, 126}

\bibitem[\protect\citeauthoryear{Islam \& Paul}{Islam \&
  Paul}{2014}]{islam2014_gx301_maxi}
Islam N.,  Paul B.,  2014, Monthly Notices of the Royal Astronomical Society,
  441, 2539

\bibitem[\protect\citeauthoryear{Islam \& Paul}{Islam \&
  Paul}{2016}]{4u1700_orbPdot_islam2016}
Islam N.,  Paul B.,  2016, Monthly Notices of the Royal Astronomical Society,
  461, 816

\bibitem[\protect\citeauthoryear{{Jahoda}, {Swank}, {Giles}, {Stark},
  {Strohmayer}, {Zhang}  \& {Morgan}}{{Jahoda} et~al.}{1996}]{rxte_jahoda}
{Jahoda} K.,  {Swank} J.~H.,  {Giles} A.~B.,  {Stark} M.~J.,  {Strohmayer} T.,
  {Zhang} W.,   {Morgan} E.~H.,  1996, in {Siegmund} O.~H.,  {Gummin} M.~A.,
  eds,  Society of Photo-Optical Instrumentation Engineers (SPIE) Conference
  Series Vol. 2808, EUV, X-Ray, and Gamma-Ray Instrumentation for Astronomy
  VII. pp 59--70, \mn@doi{10.1117/12.256034}

\bibitem[\protect\citeauthoryear{{Jenke}, {Finger}, {Wilson-Hodge}  \&
  {Camero-Arranz}}{{Jenke} et~al.}{2012}]{oao1657_orbPdot_jenke}
{Jenke} P.~A.,  {Finger} M.~H.,  {Wilson-Hodge} C.~A.,   {Camero-Arranz} A.,
  2012, \mn@doi [\apj] {10.1088/0004-637X/759/2/124}, \href
  {https://ui.adsabs.harvard.edu/abs/2012ApJ...759..124J} {759, 124}

\bibitem[\protect\citeauthoryear{{Kaper}, {Lamers}, {Ruymaekers}, {van den
  Heuvel}  \& {Zuiderwijk}}{{Kaper}
  et~al.}{1995}]{kaper_massloss_estimate_1995}
{Kaper} L.,  {Lamers} H.~J.~G.~L.~M.,  {Ruymaekers} E.,  {van den Heuvel}
  E.~P.~J.,   {Zuiderwijk} E.~J.,  1995, \mn@doi [\aap]
  {10.48550/arXiv.astro-ph/9503003}, \href
  {https://ui.adsabs.harvard.edu/abs/1995A&A...300..446K} {300, 446}

\bibitem[\protect\citeauthoryear{Klis \& Bonnet-Bidaud}{Klis \&
  Bonnet-Bidaud}{1984}]{klis1984orbital_quadratic}
Klis M.,  Bonnet-Bidaud J.,  1984, Astronomy and Astrophysics, 135, 155

\bibitem[\protect\citeauthoryear{{Koh} et~al.,}{{Koh}
  et~al.}{1997}]{koh1997-orbit}
{Koh} D.~T.,  et~al., 1997, \mn@doi [\apj] {10.1086/303929}, \href
  {https://ui.adsabs.harvard.edu/abs/1997ApJ...479..933K} {479, 933}

\bibitem[\protect\citeauthoryear{{Lai}}{{Lai}}{1996}]{donglai_psrj0045}
{Lai} D.,  1996, \mn@doi [\apjl] {10.1086/310166}, \href
  {https://ui.adsabs.harvard.edu/abs/1996ApJ...466L..35L} {466, L35}

\bibitem[\protect\citeauthoryear{{Leahy}}{{Leahy}}{1987}]{epochfold_leahy_1987}
{Leahy} D.~A.,  1987, \aap, \href
  {https://ui.adsabs.harvard.edu/abs/1987A&A...180..275L} {180, 275}

\bibitem[\protect\citeauthoryear{Leahy \& Kostka}{Leahy \&
  Kostka}{2008}]{LC_stream_model}
Leahy D.~A.,  Kostka M.,  2008, \mn@doi [Monthly Notices of the Royal
  Astronomical Society] {10.1111/j.1365-2966.2007.12754.x}, 384, 747

\bibitem[\protect\citeauthoryear{{Lecar}, {Wheeler}  \& {McKee}}{{Lecar}
  et~al.}{1976}]{lecar_tidal_circularize_herx1_cenx3}
{Lecar} M.,  {Wheeler} J.~C.,   {McKee} C.~F.,  1976, \mn@doi [\apj]
  {10.1086/154311}, \href
  {https://ui.adsabs.harvard.edu/abs/1976ApJ...205..556L} {205, 556}

\bibitem[\protect\citeauthoryear{{Levine}, {Bradt}, {Cui}, {Jernigan},
  {Morgan}, {Remillard}, {Shirey}  \& {Smith}}{{Levine}
  et~al.}{1996}]{asm_levine}
{Levine} A.~M.,  {Bradt} H.,  {Cui} W.,  {Jernigan} J.~G.,  {Morgan} E.~H.,
  {Remillard} R.,  {Shirey} R.~E.,   {Smith} D.~A.,  1996, \mn@doi [\apjl]
  {10.1086/310260}, \href
  {https://ui.adsabs.harvard.edu/abs/1996ApJ...469L..33L} {469, L33}

\bibitem[\protect\citeauthoryear{{Liu}}{{Liu}}{2020}]{liu_gx301_hardlag}
{Liu} J.,  2020, \mn@doi [\mnras] {10.1093/mnras/staa1774}, \href
  {https://ui.adsabs.harvard.edu/abs/2020MNRAS.496.3991L} {496, 3991}

\bibitem[\protect\citeauthoryear{{Lutovinov}, {Tsygankov}  \&
  {Chernyakova}}{{Lutovinov} et~al.}{2012}]{lutovinov_2012_efs_error}
{Lutovinov} A.,  {Tsygankov} S.,   {Chernyakova} M.,  2012, \mn@doi [\mnras]
  {10.1111/j.1365-2966.2012.21036.x}, \href
  {https://ui.adsabs.harvard.edu/abs/2012MNRAS.423.1978L} {423, 1978}

\bibitem[\protect\citeauthoryear{{Malacaria}, {Jenke}, {Roberts},
  {Wilson-Hodge}, {Cleveland}, {Mailyan}  \& {GBM Accreting Pulsars Program
  Team}}{{Malacaria} et~al.}{2020}]{2malacaria_gapp}
{Malacaria} C.,  {Jenke} P.,  {Roberts} O.~J.,  {Wilson-Hodge} C.~A.,
  {Cleveland} W.~H.,  {Mailyan} B.,   {GBM Accreting Pulsars Program Team}
  2020, \mn@doi [\apj] {10.3847/1538-4357/ab855c}, \href
  {https://ui.adsabs.harvard.edu/abs/2020ApJ...896...90M} {896, 90}

\bibitem[\protect\citeauthoryear{Manikantan, Paul, Roy  \& Rana}{Manikantan
  et~al.}{2023}]{manikantan_gx301_maxi}
Manikantan H.,  Paul B.,  Roy K.,   Rana V.,  2023, Monthly Notices of the
  Royal Astronomical Society, 520, 1411

\bibitem[\protect\citeauthoryear{Matsuoka et~al.,}{Matsuoka
  et~al.}{2009}]{maxi}
Matsuoka M.,  et~al., 2009, \mn@doi [Publications of the Astronomical Society
  of Japan] {10.1093/pasj/61.5.999}, 61, 999

\bibitem[\protect\citeauthoryear{{Meegan}, {Fishman}, {Wilson}, {Paciesas},
  {Pendleton}, {Horack}, {Brock}  \& {Kouveliotou}}{{Meegan}
  et~al.}{1992}]{batse_meegan}
{Meegan} C.~A.,  {Fishman} G.~J.,  {Wilson} R.~B.,  {Paciesas} W.~S.,
  {Pendleton} G.~N.,  {Horack} J.~M.,  {Brock} M.~N.,   {Kouveliotou} C.,
  1992, \mn@doi [\nat] {10.1038/355143a0}, \href
  {https://ui.adsabs.harvard.edu/abs/1992Natur.355..143M} {355, 143}

\bibitem[\protect\citeauthoryear{Mihara et~al.,}{Mihara
  et~al.}{2011}]{gsc_maxi}
Mihara T.,  et~al., 2011, \mn@doi [Publications of the Astronomical Society of
  Japan] {10.1093/pasj/63.sp3.S623}, 63, S623

\bibitem[\protect\citeauthoryear{{M{\"o}nkk{\"o}nen}, {Doroshenko},
  {Tsygankov}, {Nabizadeh}, {Abolmasov}  \& {Poutanen}}{{M{\"o}nkk{\"o}nen}
  et~al.}{2020}]{2020monkkoen}
{M{\"o}nkk{\"o}nen} J.,  {Doroshenko} V.,  {Tsygankov} S.~S.,  {Nabizadeh} A.,
  {Abolmasov} P.,   {Poutanen} J.,  2020, \mn@doi [\mnras]
  {10.1093/mnras/staa906}, \href
  {https://ui.adsabs.harvard.edu/abs/2020MNRAS.494.2178M} {494, 2178}

\bibitem[\protect\citeauthoryear{Mukherjee \& Paul}{Mukherjee \&
  Paul}{2003}]{mukherjee2003_gx301_clumpywind}
Mukherjee U.,  Paul B.,  2003, Bulletin of the Astronomical Society of India,
  31

\bibitem[\protect\citeauthoryear{{Nagase} et~al.,}{{Nagase}
  et~al.}{1982}]{nagase-intro}
{Nagase} F.,  et~al., 1982, \mn@doi [\apj] {10.1086/160551}, \href
  {https://ui.adsabs.harvard.edu/abs/1982ApJ...263..814N} {263, 814}

\bibitem[\protect\citeauthoryear{Naik \& Paul}{Naik \&
  Paul}{2004}]{lmcx4_orbPdot_Naik_2004}
Naik S.,  Paul B.,  2004, \mn@doi [The Astrophysical Journal] {10.1086/379803},
  600, 351

\bibitem[\protect\citeauthoryear{{Parkes}, {Mason}, {Murdin}  \&
  {Culhane}}{{Parkes} et~al.}{1980}]{parkes_gx301_windvelo_windrate_1980}
{Parkes} G.~E.,  {Mason} K.~O.,  {Murdin} P.~G.,   {Culhane} J.~L.,  1980,
  \mn@doi [\mnras] {10.1093/mnras/191.3.547}, \href
  {https://ui.adsabs.harvard.edu/abs/1980MNRAS.191..547P} {191, 547}

\bibitem[\protect\citeauthoryear{{Paul}}{{Paul}}{2017}]{paul_hmxb_orb_evolve_review}
{Paul} B.,  2017, \mn@doi [Journal of Astrophysics and Astronomy]
  {10.1007/s12036-017-9475-4}, \href
  {https://ui.adsabs.harvard.edu/abs/2017JApA...38...39P} {38, 39}

\bibitem[\protect\citeauthoryear{Paul \& Naik}{Paul \&
  Naik}{2011}]{paul2011transient}
Paul B.,  Naik S.,  2011, arXiv preprint arXiv:1110.4446

\bibitem[\protect\citeauthoryear{Pravdo \& Ghosh}{Pravdo \&
  Ghosh}{2001}]{Pravdo_2001}
Pravdo S.~H.,  Ghosh P.,  2001, \mn@doi [The Astrophysical Journal]
  {10.1086/321350}, 554, 383

\bibitem[\protect\citeauthoryear{Raichur \& Paul}{Raichur \&
  Paul}{2010}]{cenx3_orbPdot_raichur_rxte}
Raichur H.,  Paul B.,  2010, \mn@doi [Monthly Notices of the Royal Astronomical
  Society] {10.1111/j.1365-2966.2009.15778.x}, 401, 1532

\bibitem[\protect\citeauthoryear{{Raman}, {Varun}, {Paul}  \&
  {Bhattacharya}}{{Raman} et~al.}{2021}]{graman_efs_error}
{Raman} G.,  {Varun} B.,  {Paul} B.,   {Bhattacharya} D.,  2021, \mn@doi
  [\mnras] {10.1093/mnras/stab2835}, \href
  {https://ui.adsabs.harvard.edu/abs/2021MNRAS.508.5578R} {508, 5578}

\bibitem[\protect\citeauthoryear{{Sato}, {Nagase}, {Kawai}, {Kelley},
  {Rappaport}  \& {White}}{{Sato} et~al.}{1986}]{sato1986-orb}
{Sato} N.,  {Nagase} F.,  {Kawai} N.,  {Kelley} R.~L.,  {Rappaport} S.,
  {White} N.~E.,  1986, \mn@doi [\apj] {10.1086/164157}, \href
  {https://ui.adsabs.harvard.edu/abs/1986ApJ...304..241S} {304, 241}

\bibitem[\protect\citeauthoryear{{Shirke}, {Bala}, {Roy}  \&
  {Bhattacharya}}{{Shirke} et~al.}{2021}]{cenx3_orbitalP_astrosat}
{Shirke} P.,  {Bala} S.,  {Roy} J.,   {Bhattacharya} D.,  2021, \mn@doi
  [Journal of Astrophysics and Astronomy] {10.1007/s12036-021-09710-w}, \href
  {https://ui.adsabs.harvard.edu/abs/2021JApA...42...58S} {42, 58}

\bibitem[\protect\citeauthoryear{{Singh}, {Naik}, {Paul}, {Agrawal}, {Rao}  \&
  {Singh}}{{Singh} et~al.}{2002}]{singh_cygx3}
{Singh} N.~S.,  {Naik} S.,  {Paul} B.,  {Agrawal} P.~C.,  {Rao} A.~R.,
  {Singh} K.~Y.,  2002, \mn@doi [\aap]
  {10.1051/0004-6361:2002092310.48550/arXiv.astro-ph/0206134}, \href
  {https://ui.adsabs.harvard.edu/abs/2002A&A...392..161S} {392, 161}

\bibitem[\protect\citeauthoryear{Tauris \& Van Den~Heuvel}{Tauris \& Van
  Den~Heuvel}{2006}]{tauris_vandenheuvel_2006}
Tauris T.,  Van Den~Heuvel E.,  2006, Formation and evolution of compact
  stellar X-ray sources.
pp 623--665

\bibitem[\protect\citeauthoryear{Thorne \& Zytkow}{Thorne \&
  Zytkow}{1977}]{thorne1977stars}
Thorne K.~S.,  Zytkow A.,  1977, The Astrophysical Journal, 212, 832

\bibitem[\protect\citeauthoryear{{Witte} \& {Savonije}}{{Witte} \&
  {Savonije}}{1999}]{witte_tide}
{Witte} M.~G.,  {Savonije} G.~J.,  1999, \mn@doi [\aap]
  {10.48550/arXiv.astro-ph/9909073}, \href
  {https://ui.adsabs.harvard.edu/abs/1999A&A...350..129W} {350, 129}

\bibitem[\protect\citeauthoryear{{van den Heuvel}}{{van den
  Heuvel}}{1994}]{interactingbinaries_vandenheuvel}
{van den Heuvel} E.~P.~J.,  1994, in Saas-Fee Advanced Course 22: Interacting
  Binaries. pp 263--474

\makeatother
\end{thebibliography}

%\begin{thebibliography}{99}
%\end{thebibliography}

%%%%%%%%%%%%%%%%%%%%%%%%%%%%%%%%%%%%%%%%%%%%%%%%%%

%%%%%%%%%%%%%%%%% APPENDICES %%%%%%%%%%%%%%%%%%%%%

\appendix

\section{Error estimation by Bootstrap}\label{appendix:errorefsP}
\begin{itemize}
    \item In each of the dwell lightcurves used, count-rate in the $i$\textsuperscript{th} temporal bin $c_i$ was replaced with $c_i + x\sigma_i$, where $x$ is independently randomly sampled from the uniform distribution $\mathcal{U}(-1,1)$ (See \citealt{lutovinov_2012_efs_error}, \citealt{boldin_2013_efs_error} and \citealt{graman_efs_error}).
    \item Using this technique, 1000 sample lightcurves were simulated for each long-term lightcurve and pulsed flux history.
    \item The best period from each simulated lightcurve was estimated by fitting a \texttt{gaussian} to the $\chi^2$ vs $P$\textsubscript{orb} plot and retrieving the best-fit \texttt{gaussian} centre. 
  \item The mean ($\mu$) and standard deviation ($\sigma$) of the distribution of best-fit \texttt{gaussian} centres for 1000 simulations from each lightcurve were assigned its $P$\textsubscript{orb} and $\Delta P$\textsubscript{orb}, respectively.
\end{itemize}

\section{Energy dependence of flares}\label{appendix:flareEdep}
To assess the energy dependence of the arrival time of pre-periastron flares, we used the overlapping duration of \textit{Swift}/BAT (15--50 keV) lightcurve with \textit{RXTE}/ASM (1.5--12 keV), \textit{MAXI} (2--20 keV, 2--4 keV, 4--10 keV, 10--20 keV) and \textit{Fermi}/GBM (12--50 keV) (Fig.~\ref{fig:ltlc}), and the overlapping duration of \textit{RXTE}/ASM with \textit{CGRO}/BATSE. BAT and BATSE were selected as reference lightcurves, and the below steps were performed individually for both. 
\begin{itemize}
    \item The overlapping duration between lc\textsubscript{\textit{ref}} and each lc\textsubscript{\textit{oth}}s were first identified, where lc\textsubscript{\textit{ref}} is the refernce lightcurve (BAT or BATSE) and lc\textsubscript{\textit{oth}} is the other lightcurve having an overlap with lc\textsubscript{\textit{ref}}.
    \item {\small XRONOS} compatible window files were created using the {\small HEASOFT} tool \textit{xronwin} to restrict data to the overlap duration.
    \item In the overlap duration, lc\textsubscript{\textit{ref}} and lc\textsubscript{\textit{oth}} were folded at an arbitrary reference epoch ($T_\textrm{fold}$) with the average of the orbital periods ($P_\textrm{orb}$) derived from the two lightcurves (Table~\ref{tab:ltlc-efsearch}).
    \item The vicinity of the flare in each folded orbital intensity profile was modelled with a \texttt{constant+lorentzian} and the centre of \texttt{lorentzian} was estimated along with its $2.7\sigma$ error. The centre of \texttt{lorentzian} is assigned as the phase of flare peak ($\phi_\textrm{flare}$).
    \item Number of orbits elapsed since $T_\textrm{fold}$ to the middle of each window was estimated by \textit{floor}((T\textsubscript{window-mid}$-$48370.5)/P\textsubscript{orb}). Flare time for \textit{i}\textsuperscript{th} lightcurve was estimated by $T_\textrm{flare,i}=T_\textrm{fold}+NP_\textrm{orb,i}+P_\textrm{orb,i}\phi_\textrm{peak,i}$ .
    \item The delay ($\Delta T$\textsubscript{flare}) between the flare times of the reference lightcurve and the other lightcurve was calculated (Table~\ref{tab:overlap_lc_2}).
\end{itemize}

\section{Orbital period derivative from timing signature of the pre-periastron flares}\label{appx:timingflarepeak}

The pulsed histories from BATSE (20--50 keV) and GBM (12--50 keV), and the long-term lightcuves from ASM (1.5--12 keV), BAT (15--50 keV), and MAXI (10--20 keV) were used to estimate the time signature of flare peaks. The steps were followed in the order in which they are listed below: 

\begin{itemize}
    \item Each lightcurve was split into three equal slices (windows) and is folded with the respective orbital period (Table~\ref{tab:ltlc-efsearch}) at the epoch corresponding to the beginning of the window. The idea is to find three representative flare-peak times per lightcurve.
    \item The maximum SNR for orbital intensity profile was obtained from BAT, where it was also found that \texttt{lorentzian} is a better fit to the flare compared to \texttt{gaussian} based on weighted variance. Therefore, a \texttt{constant+lorentzian} was fit on the folded lightcurve in the vicinity of flare ($\phi$\textsubscript{orb}), and the centre of the best fit \texttt{lorentzian} was assigned the phase of flare peak $\phi_\textrm{flare}\pm\Delta\phi_\textrm{flare}$.
    \item Orbital cycles $n$ elapsed since the start of the window to the flare peak ($T$\textsubscript{flare}) in each slice of the lightcurve was estimated by \textit{floor}((T\textsubscript{slice-mid}$-$48370.5)/P\textsubscript{orb}). The flare time was then estimated by $T_\textrm{flare}=T_\textrm{win-begin} + nP_\textrm{orb} + \phi_\textrm{peak}P_\textrm{orb}$.
    \item The data $T$\textsubscript{flare} vs $n$ was fitted with a linear function, and the residuals to the best fit linear function were then checked for a quadratic trend indicative of orbital evolution. 
\end{itemize}

\section{Tidal evolution}\label{Appx:tidalcirctimescale}

Under the weak friction model approximation, due to the internal frictional properties of the companion, the formation of a tidal bulge occurs $\tau$ s after the compact object exerts gravitational force to raise it. $\tau$ is called the tidal time lag and by this duration, the compact object would have moved a relative angular displacement of $\delta=\tau\sigma$ about the tidal bulge, called the tidal lag angle. This displacement of the tidal bulge relative to the line connecting two stars results in a tidal torque that affects the binary orbit. The tidal time lag ($\tau$) is related to the properties of stellar structure. The degree of response of the binary orbit to the tidal forces is represented by the apsidal motion constant $k$. $\sigma\approx\Omega_c-\Omega_\textrm{orb}$ is the apparent angular velocity of NS relative to the surface of the companion.

The rate of change of the semi-major axis due to tidal circularization of the binary is given by \cite{tidal_hut1981} as the following equation~\ref{eq:hut}
\begin{equation}\label{eq:hut}
    \frac{\dot a}{a} = \frac{2\dot P_\textrm{orb}}{3P_\textrm{orb}} = -6\frac{k}{T}\grave{q}(1+\grave{q})\left(\frac{R_c}{a}\right)^8\frac{1}{(1-e^2)^{7.5}}\left[f_1(e^2)-(1-e^2)^{1.5}f_2(e^2)\frac{\Omega_c}{n}\right]
\end{equation}
where,\newline
$a$ and $P$\textsubscript{orb} are the semi-major axis and orbital period, and $\dot a$ and $\dot P$\textsubscript{orb} their rate of changes, $R_c$ is the companion radius, $e$ is the binary eccentricity, $\Omega_c$ is the rotation frequency of the companion. After substituting the known parameters of GX 301--2 from (Table~\ref{tab:knownpars}),
\begin{equation*}
    f_1(e^2) = 1+\frac{31}{2}e^2+\frac{255}{8}e^4+\frac{185}{16}e^6+\frac{25}{64}e^8 \sim6.1
\end{equation*}
\begin{equation*}
    f_2(e^2) = 1+\frac{15}{2}e^2+\frac{45}{8}e^4+\frac{5}{16}e^6 \sim2.9
\end{equation*}
\begin{equation*}
    n = \sqrt{\frac{G(M_x+M_c)}{a_x^3}} \sim 1.9\times10^{-6}\ \textrm{rad\ s}^{-1}
\end{equation*}
\begin{equation*}
    \Omega_c = \frac{2\pi}{P_c} \sim 1.01\times10^{-6}\  \textrm{rad\ s}^{-1}
\end{equation*}
\begin{equation*}
    \grave{q} = \frac{M_x}{M_c} \sim0.03
\end{equation*}
\begin{equation*}
    T = \frac{R_c^3}{GM_c\tau} \sim \frac{3.3\times10^{10}}{\tau}\ \textrm{s}
\end{equation*}
Substituting in equation~\ref{eq:hut}

\begin{equation*}
    -\frac{2\times5.52\times10^{-13}}{3} \approx -6\frac{k\tau}{3.3\times10^{10}}\times0.029\times0.003\times6.5\times5.05
\end{equation*}
\begin{equation*}
    k\tau \approx 0.64\ \textrm{s}
\end{equation*}
If tidal dissipation is assumed to be facilitated by an outer convective layer around the stellar core of Wray 15-977, equation A1 in \cite{lecar_tidal_circularize_herx1_cenx3} gives the relation of $k\tau$ to the characteristics of such a convection layer as
\begin{equation}\label{eq:lecar}
    k\tau \approx 25\ s\frac{\lambda\eta v_\textrm{conv}(\textrm{km\ s}^{-1})}{(g/g_\odot)}
\end{equation}
where, $\lambda$ (fractional depth of convective layer), $\eta$ (fractional mass of the convective zone) and $v$\textsubscript{conv} (convective velocity) define the property of the convective envelope, and
\begin{equation*}
    \frac{g}{g_\odot} = \frac{(M_c/\mathrm{M}_\odot)}{(R_c/\mathrm{R}_\odot)^2} \approx 0.007
\end{equation*}
Substituting $k\tau=0.71$ s in equation~\ref{eq:lecar} gives
\begin{equation*}
    \lambda\eta v_\textrm{conv} \approx 1.68\times10^{-4}\ \textrm{km\ s}^{-1}
\end{equation*}

\section{O--C curve without correcting for energy dependence of flares} \label{Appx:rawflare_orbPdot}

\begin{figure}
    \centering
    \includegraphics[width=\columnwidth]{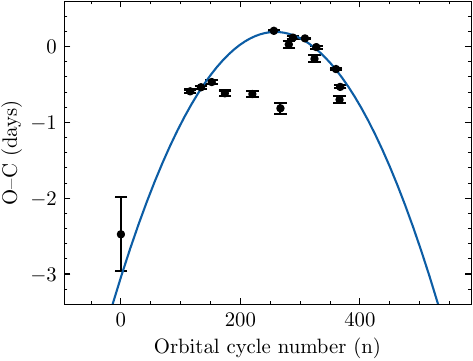}
    \caption{Results of fitting a quadratic function on the O--C curve that was generated without correcting for the energy dependence of the pre-periastron flares. The fit returned a best fit $\dot P_\mathrm{orb}=-2.3\times10^{-6}$ s s$^{-1}$, albeit with a very large fit statistic. }
    \label{fig:appx_o-c_raw}
\end{figure}

When the energy dependence of the pre-periastron flares were not taken into account for generating the O--C curve as opposed to the analysis described in Section~\ref{subsec:o-c}, a quadratic fit on the resulting O--C curve yielded a best fit $\dot P_\mathrm{orb}=-2.3\times10^{-6}$ s s$^{-1}$ with a very large \texttt{wvar} of 965 for 13 d.o.f  (Fig.~\ref{fig:appx_o-c_raw}).
%%%%%%%%%%%%%%%%%%%%%%%%%%%%%%%%%%%%%%%%%%%%%%%%%%

% Don't change these lines
\bsp	% typesetting comment
\label{lastpage}
\end{document}